\documentclass[proof]{WileyASNA-v1_MC}

\articletype{ORIGINAL ARTICLE}

\received{\today}
\revised{}
\accepted{}

\usepackage{savesym}

\usepackage[rightcaption]{sidecap}
\usepackage{nicefrac}

\usepackage[title]{appendix}
\usepackage{amsmath}
\usepackage[export]{adjustbox}
\usepackage{graphicx}
\usepackage{txfonts}
\usepackage{textcomp}
\usepackage[flushleft]{threeparttable}
\usepackage[utf8]{inputenc}
\usepackage{natbib}
\usepackage{lscape}
\bibpunct{(}{)}{;}{a}{}{,}

\usepackage{amsfonts}
\usepackage{color}
\usepackage{graphics}
\graphicspath{}
\usepackage{multirow}

\usepackage{geometry}
\usepackage{amssymb}
\usepackage{mathtools}
\usepackage{graphicx}
\usepackage{epstopdf}
\usepackage{pstricks}
\usepackage{pst-plot}
\usepackage{pstricks-add}
\usepackage{epsf}
\usepackage{pgf}
\usepackage{eqnarray,amsmath}

\usepackage{hyperref}

\newcommand{\ns}{\hspace*{-5pt}}

\definecolor{indiagreen}{rgb}{0.07, 0.53, 0.03}

\begin{document} 

%===============================================================================
%   TITLE, AUTHORS, AND AFFILIATIONS
%===============================================================================
\title{On the parameter refinement of inflated exoplanets with large radius uncertainty based on TESS observations}

\author[1,2,3]{X.\ Alexoudi*}

\authormark{X.\ Alexoudi \textsc{et al.}}

\address[1]{\orgname{Leibniz-Institut f\"{u}r Astrophysik Potsdam (AIP)}, 
    \orgaddress{\city{Potsdam}, 
    \country{Germany}}}

\address[2]{\orgname{Universit{\"a}t Potsdam},
    \orgdiv{Institut f{\"u}r Physik und Astronomie}, 
    \orgaddress{\city{Potsdam}, 
    \country{Germany}}}

\address[3]{\orgname{Potsdam Graduate School}, 
    \orgaddress{\city{Potsdam}, 
    \country{Germany}}}

\corres{*\email{xalexoudi@aip.de}}

%===============================================================================
%   ABSTRACT AND KEYWORDS
%===============================================================================

\abstract{We revisited ten known exoplanetary systems using publicly available data provided by the Transiting Exoplanet Survey Satellite (TESS). The sample presented in this work consists of short period transiting exoplanets, with inflated radii and large reported uncertainty on their planetary radii. The precise determination of these values is crucial in order to develop accurate evolutionary models and understand the inflation mechanisms of these systems. 
Aiming to evaluate the planetary radius measurement, we made use of the planet-to-star radii ratio, a quantity that can be measured during a transit event. We fit the obtained transit light curves of each target with a detrending model and a transit model. Furthermore, we used emcee, which is based on a Markov chain Monte Carlo approach, to assess the best fit posterior distributions of each system parameter of interest. We refined the planetary radius of \mbox{WASP-140 b} by approximately $12\%$, and we derived a better precision on its reported asymmetric radius uncertainty by approximately $86\%$ and $67\%$. We also refined the orbital parameters of \mbox{WASP-120 b} by $2\sigma$. Moreover, using the high-cadence TESS datasets, we were able to solve a discrepancy in the literature, regarding the planetary radius of the exoplanet \mbox{WASP-93 b}. For all the other exoplanets in our sample, even though there is a tentative trend that planetary radii of (near-) grazing systems have been slightly overestimated in the literature, the planetary radius estimation and the orbital parameters were confirmed with independent observations from space, showing that TESS and ground-based observations are overall in good agreement.
}
\keywords{ techniques: photometric, stars: WASP-140, WASP-136, WASP-113, WASP-120, WASP-93, HAT-P-16, WASP-123, WASP-76, WASP-20, WASP-108, planetary systems }

\jnlcitation{\cname{%
    \author{X.\ Alexoudi}}
     (\cyear{2021}), 
\ctitle{On the parameter refinement of inflated exoplanets with large radius uncertainty based on TESS observations}, 
\cjournal{ASNA}, \cvol{}.}

\maketitle

\section{Introduction}

The field of exoplanets is a rapidly advancing domain in modern astrophysics. Surveys and missions were dedicated through the years with joint ground- and space-based efforts in the discovery of exoplanets and the characterization of their interiors (e.g. The Hungarian Automated Telescope Network (HATNet) project \citep{Bakos2018}, the SuperWASP: Wide Angle Search for Planets project \citep{Street2003}, the CoRoT project \citep{Barge2008} and the Hubble Space Telescope (HST)). 
The results of these endeavors has shown that exoplanets are not similar to the planets of our solar system; increasing this way the interest of the scientific community to investigate further these unknown exotic worlds. For example, the Kepler mission \citep{Borucki2010} (and later on the K2 mission \citep{Howell2014}, a follow-on to the Kepler mission), provided thousands of transiting systems, where the planet orbits its host star on an edge-on orbit, as seen by an observer on the Earth. Those systems included Earth-like planets, Neptune-sized, and interestingly large gaseous planets of the size of Jupiter at short orbital periods. However, most of Kepler's targets are faint stars, and the atmospheric characterization of large gaseous transiting exoplanets is favorable only for the brightest candidates. 

One mission that is focused specifically on bright targets ($5\%$ on brighter than $V_{mag}=8$) is the Transiting Exoplanet Survey Satellite (TESS), which was launched in 2018 \citep{Ricker2015} and it is scheduled to fixate its detectors on more than 100 exoplanets. For more than three years, TESS has been observing the night sky and has been providing datasets of photometric monitoring of bright stars and their planets, triggering follow-up studies by ground-based facilities and setting up the challenge for future space-born missions (e.g. the James Webb Space Telescope, JWST, of \cite{Gardner2006}). 

The radii of inflated hot Jupiters, strongly affected by the irradiation of their host stars, extend beyond the typical Jupiter radius size. They can be determined from uninterrupted high-quality photometric light curves, always with respect to their host star radius measurement. The dimming in the brightness of these stars is larger for transiting gas giants with extended inflated radii. Even though, the planet radius estimation is derived straightforward from the planet-to-star radii ratio measurement during a transit event, the uncertainty on this measurement can be constrained significantly due to different factors; from unknown systematic errors to incomplete datasets, or a different approach on the methodology that was used for the transit light curve analysis. The required precision for the planetary radius measurement can contribute to the general understanding of the inflation scenarios taking part on different exoplanets. 

Some of the most prominent mechanisms able to trigger the effect of inflation on gas giant exoplanets are: the irradiating flux sourcing from the host star itself, that heats up the planet and increases its equilibrium temperature \citep{GuillotShowman2002}, the ohmic heating mechanism \citep{Laughlin2011}, as a result from the coupling of the atmospheric flows with the magnetic field of the planet, the kinetic heating (a more direct mechanism), as some incident flux turns into kinetic energy and eventually into thermal energy that heats up the atmosphere. Last but not least, another mechanism is the tidal heating promoted by the circularization of the planetary orbit \citep{Bodenheimer2001,Leconte2010}. Evolutionary models that predict the formation mechanisms of inflated exoplanets can gain in robustness with the precise measurements of the key physical and orbital parameters of those exotic systems, and conclude to a suitable explanation for the applied inflation mechanism.  

Moreover, studies on transiting exoplanets can yield an accurate planetary radius, that in combination with high precision radial velocity (RV) observations, can provide a mass for the planet, hence a mean density estimation which gives important information regarding the internal structure of these planets. Furthermore, the precise radius estimation and the distance from their hosts, can give insights on the gravitational potential of the planets and their equilibrium temperature. Consequently, those computations are useful in transmission spectroscopy because they provide an estimation of the extension of an exoplanetary atmosphere, if it is present. The correct characterization of this, is based on the investigation of the planetary to stellar radii ratio over different wavelengths of observation, known as the transmission spectrum. The employment of incorrect parameters in the light curve analysis can compromise the structure of the spectrum and yield misplaced slopes \citep{Alexoudi2020}. TESS is expected to contribute to those studies by providing precise physical and orbital parameters derived from high-quality high-cadence datasets.  

Motivated by the aforementioned capabilities of TESS, we obtained a sample of inflated hot giants, orbiting bright stars in short close-in orbits, and proceeded with a parameter refinement of those systems since their parameters have not been up-to-date for more than three years (see Table~\ref{sample_params}{\ns}). The dates of the last update are presented as registered at the NASA Exoplanet Archive\footnote{\href{http://exoplanetarchive.ipac.caltech.edu}{exoplanetarchive.ipac.caltech.edu}}. In this work, we focused on exoplanets with the largest reported planetary radius uncertainty, and we expected TESS to ameliorate our knowledge on the planetary radii of those systems and their physical properties. The aim is to provide the most accurate parameterization of these systems and quantify TESS's capabilities in comparison to the ground-based facilities. 

The structure of this paper is the following: Section 2 presents the TESS observations of each exoplanet of our sample. In Section 3, we describe the reduction method that was employed for the analysis of these datasets and in Section 4, we demonstrate the adopted methods in order to derive the system parameters of our targets. In Section 5, we present our results and in Section 6 we discuss the impact of these findings regarding the characterization of hot giant exoplanets. In the end, in Section 7, we provide a summary and the conclusions of this entire work. 

\begin{table*}
\begin{center}
%\begin{tabular}{lccccccc}
\begin{tabular}{cccccc}
\hline
\noalign{\smallskip}
    Exoplanet &TESS mag &P (d) & $\mathrm{R_p (R_J)}$ & Date of last update &Publication \\
\hline
\noalign{\smallskip}
WASP-140 b &$10.3$	&$2.2$	&$1.440  \substack{+0.42	\\-0.18}$& 	2016-11-30 & \cite{Hellier2017}\\
WASP-136 b &$9.5$	&$5.2$	&$1.380	\pm{0.160}$  & 2016-11-30 & \cite{Lam2017}\\
WASP-113 b&$11.2$	&$4.5$	&$1.409	\substack{+0.096\\-0.140}$&2016-07-14 &\cite{Barros2016}\\
WASP-120 b&$10.6$	&$3.6$	&$1.473	\pm{0.096}$ &2016-06-01 &\cite{Turner2016}\\
WASP-93 b &$10.6$	&$2.7$	&$1.597	\pm{0.077}$  & 2016-09-06 & \cite{Hay2016}\\
HAT-P-16 b&$10.8$	&$2.8$	&$1.289	\pm{0.066}$ &2014-05-14 &\cite{Buchhave2010}\\
WASP-123 b&$10.4$	&$3.0$	&$1.318	\pm{0.065}$&2016-06-01 &\cite{Turner2016}\\
WASP-76 b &$9.0$	&$1.8$	&$1.830	\substack{+0.060 \\	-0.040}$ &2016-01-20 &\cite{West2016}\\
WASP-20 b &$10.7$  & $4.9$ &$1.459 \pm{0.057}$   & 2015-03-05 &\cite{Anderson2015} \\
WASP-108 b &$11.2$ &$2.7$  &$1.215 \pm{0.04}$ & 2014-10-29 &\cite{Anderson2014} \\
\hline 
\end{tabular}
\caption{The sample of this work. The sample selection is based on exoplanets with
inflated radii $\mathrm{(R_p > 1.2 R_J)}$, that orbit relatively
bright stars (TESS mag < 12), in short orbital periods (P < 5 days). The
planetary radius of each target has last been updated between the years 2014 and
2017, and therefore its refinement is necessary. The targets are sorted by the uncertainty on $\mathrm{R_p}$ with a decreasing order.} 
\label{sample_params}
\end{center}
\end{table*} 

\section{Observations}
Our sample consists solely of TESS observations, spanning the period of 2018-2020. We re-visited ten inflated hot giants (nine hot Jupiters and one Saturn-sized planet) using the publicly available, two-minute cadence data of TESS. The complete list of the observations is presented in Table \ref{tess_observations}{\ns}. The obtained light curves were processed by the Science Processing Operations Center (SPOC) pipeline, based on the work of \cite{Jenkins2016}.  We made use of the PDCSAP\_FLUX, which is the Pre-search Data Conditioning SAP flux, in order to access SPOC's data. PDCSAP\_FLUX has an advantage over the simple aperture photometry (SAP\_FLUX), because of the use of the Cotrending Basis Vectors (CBVs). CBVs remove longstanding systematic trends and provide better data quality \citep{TenenbaumJenkins2018}. Another approach would be to use the light curves derived from the data validation timeseries (DVT) files as in \cite{RiddenHarper2020}, however a standard process for the analysis of TESS data is the use of the PDC light curve \citep{Shporer2019,Espinoza2020,Turner2021}. With the use of PDCSAP products, we obtained light curves corrected for pointing and focus related systematics, for the cosmic rays' contribution to the detector, for persistent outliers and flux contamination \citep{Jenkins2016}. Moreover, we performed a further selection criterion on the datasets by masking out the bad cadences. During the observations some pixels may be contaminated by various effects e.g. spacecraft is in coarse point, reaction wheel desaturation event, cosmic ray detected, stray light from the Earth or Moon (see a complete set of such effects in Table 32 in \cite{TenenbaumJenkins2018}). To account for this, we used a conservative setting (in the lightkurve package - \cite{LightkurveCollaboration2018}) in our analysis that excludes cadences with data-quality issues \citep{Littlefield2019}. 
All the cadences considered in this work are of high-quality, are products of the SPOC pipeline, and publicly available at the Mikulski Archive for Space Telescopes (MAST) \footnote{https://archive.stsci.edu/}. 

\begin{table*}
\begin{center}
%\begin{tabular}{lccccccc}
\begin{tabular}{ccccccc}
\hline
\noalign{\smallskip}
    &Target &Sector &Start Date &End Date &Cycle & Camera \\
\hline
\noalign{\smallskip}
\hline
\noalign{\smallskip}
&WASP-140 b &4 &$2018-Oct-18$ &$2018-Nov-15$ &$1$ &$2$\\ 
& &5 &$2018-Nov-15$ &$2018-Dec-11$ &$1$ &$2$\\
& &31 &$2020-Oct-21$ &$2020-Nov-19$ &$3$ &$2$\\
\hline 
&WASP-136 b &29 &$2020-Aug-26$ &$2020-Sept-22$ &$3$ &$1$\\ 
&& 42 &$2021-Aug-20$ &$ 2021-Sep-16$ &$4$ &$2$\\
\hline
&WASP-113 b &23 &$2020-Mar-18$ &$2020-Apr-16$ &$2$ &$3$\\
&  &24 &$2020-Apr-16$ &$2020-May-13$ &$2$ &$2$\\
\hline
&WASP-120 b &4 &$2018-Oct-18$ &$2018-Nov-15$ &$1$ &$3$\\
&  &5 &$2018-Nov-15$ &$2018-Dec-11$ &$1$ &$3$\\
&  &30 &$2020-Sep-22$ &$2020-Oct-21$ &$3$ &$3$\\
&  &31 &$2020-Oct-21$ &$ 2020-Nov-19$  &$3$ &$3$\\
\hline
&WASP-93 b &17 &$2019-Oct-07$ &$2019-Nov-02$ &$2$ &$2$\\ 
\hline
&HAT-P-16 b &17 &$2019-Oct-07$ & $2019-Nov-02$  &$2$ &$2$\\
\hline
&WASP-123 b &13 &$2019-Jun-19$ & $2019-Jul-18$  &$1$ &$1$\\
&           &27 &$2020-Jul-04$ &$2020-Jul-30$&$3$ &$1$\\
\hline
&WASP-76 b &30 &$2020-Sep-22$&$2020-Oct-21$ &$ 3$&$1$\\
&& 42 &$2021-Aug-20$ &$ 2021-Sep-16$ &$4$ &$3$\\
\hline
&WASP-20 b &2 & $ 2018-Aug-22 $ &$2018-Sep-20$ &$ 1$&$1$\\
& &29 &$2020-Aug-26 $ &$2020-Sep-22$&$ 3$&$1$\\
\hline
&WASP-108 b &11 &$2019-Apr-22$ & $2019-May-21$ &$ 1$&$2$\\
& & 37 & $2021-Apr-02$ &$2021-Apr-28$ &$3$&$2$\\
&&  38 & $2021-Apr-28$ &$2021-May-26$ &$3$&$2$\\
\hline
\hline 
\end{tabular}
\caption{The TESS observations of each target, that were used in this work. We provide information on the sector, on the date, the observing cycle of each observation and the camera that was used. For many of the targets, there were available datasets of observations from multiple sectors.} 
\label{tess_observations}
\end{center}
\end{table*}

\section{Data analysis}
We made use of Lightkurve \citep{LightkurveCollaboration2018}, a Python package for Kepler and TESS data analysis. We cleaned additionally the light curves from outliers to the $6\sigma$ level, and normalize them by the median. We applied a further correction in the light curves by removing additional trends using the flatten method of the lightkurve package. This correction removes long-term trends using a Savitzky-Golay filter. We applied a window length of the filter of 1501 points and a break tolerance (in order to account for any large gaps in time) of 50. For each individual light curve of each sector, we applied a second order time-dependent polynomial, aiming to remove any remaining trends. We made use of the Bayesian Information Criterion, BIC \citep{Schwarz1978}, to determine the best detrending model for our transit light curve fitting \citep{Mallonn2015, Mallonn2016}, and we concluded to a second order time-dependent polynomial that yields a smaller BIC value. In the end, we folded the light curves to a common transit mid-time reference of zero. Then, we used a combination of a detrending polynomial for the folded light curve and the Bad-Ass Transit Model cAlculatioN (BATMAN software by \citet{Kreidberg2015}) in order to fit our data.  We adopted initial model parameters as defined for each exoplanet from their discovery papers and used the Barycentric Julian Date as the Barycentric Dynamical Time ($\mathrm{BJD_{TDB}}$) standard for the mid-time, as it is generally recommended being used in practice for astrophysical events \citep{Eastman2010}. We made use of the quadratic limb darkening law \citep{Howarth2011}, and employed coefficients from the limb darkening calculator of the Exoplanet Characterization Toolkit\footnote{\href{https://exoctk.stsci.edu/limb\_darkening}{exoctk.stsci.edu/limb\_darkening}}. We preferred limb darkening coefficients (LDCs) calculated with the ATLAS stellar atmospheric model grids, because there is an offset between the theoretical and the observed TESS LDCs when using the PHOENIX models, while using the ATLAS models there is a significantly smaller offset \citep{Claret2017}. Then, we chose the traditional Cousins I - band, for the wavelength band for which to obtain the LDCs, because the TESS detector bandpass is centered on 786.5~nm\footnote{\href{https://heasarc.gsfc.nasa.gov/docs/tess/the-tess-space-telescope.html}{heasarc.gsfc.nasa.gov/docs/tess/the-tess-space-telescope.html}}. And finally, we proceeded with the light curve fit process. That being the case, we kept all the parameters fixed to their theoretical values, except the time of the mid-transit $T_0$, the orbital inclination $i$, the semi major axis normalized in stellar radii $a/Rs$, the ratio of the planet to star radii $R_p/R_s$, and the three terms of the time polynomial $c_0,c_1$ and $c_2$. The orbital period P, the eccentricity $e$ and the limb darkening coefficients remained fixed to their theoretical values during the fitting process. The free parameters were fit through the maximum likelihood optimization. For this purpose, we used the "optimize" module from SciPy \footnote{\href{https://docs.scipy.org/doc/scipy/reference/optimize.html}{https://docs.scipy.org/doc/scipy/reference/optimize.html}}, in order to apply a numerical optimization to the likelihood function and derive the parameters that maximize it.
We made use of the maximum likelihood estimation of the free model parameters and employed the emcee \footnote{\href{https://emcee.readthedocs.io/en/stable/}{emcee.readthedocs.io/en/stable/}} approach \citep{ForemanMackey2013} to fit the combined transit model on the data and obtain posterior distributions for each parameter with errors.  
We used uniform prior values from where the emcee can draw samples in order to define the posterior values. The final probability function is a sum of the prior function and the likelihood function. We set the initialization of 30 walkers around the maximum likelihood estimations of each parameter and then run 20000 iterations. We access the samples using the ``EnsembleSampler.get chain'' method and identify the parameter values for each walker and for each iteration of the chain. The walkers initially wonder around the maximum likelihood values of each parameter, and then very quickly start to converge towards the full posterior distribution. There is a burnt-in phase of around 10000 steps. We ensured convergence of the chains with the integrated autocorrelation time $\tau$ metric, which computes the autocorrelation time of the emcee. Usually, chains longer than $50 \times \tau$ are sufficient and the burnt-in phase of 10000 steps ensured convergence for all the chains of the analyses of all our targets.  Then, we thinned each chain and flatten it, so we would obtain a final flat list of samples. We present the best-fit parameters for each target in corner plots, where we can see the projections of the posterior probability distributions of our parameters. The 2-D histograms show the marginalized distribution of each. We used the uncertainties based on the $16^{th}$, $50^{th}$ and, $84^{th}$ percentiles of the samples. These confidence intervals correspond to $\pm 1 \sigma$ for a Gaussian posterior distribution.
The best fit of the modeled transit light curves and the corner plots of the analysis of each exoplanet of our sample are presented in the Appendix A. 

\section{Derivation of the physical parameters} 
The investigation of inflated giant exoplanets with high-quality photometric TESS observations allows for a more concrete estimate of the properties of those systems. In our analysis, we adopted the values for the stellar radius $\mathrm{R_s}$, the $e$ and the RV semi-amplitude of the stars $\mathrm{K_\star}$, of these systems from their discovery papers. The newly derived parameters with TESS: $T_0$, $R_p/R_s$, $a/R_s$, $i$, can provide a refined measurement on their planetary surface gravitational acceleration, $\mathrm{g_p}$ and on their equilibrium temperature, $\mathrm{T_{eq}}$. The updated $\mathrm{R_p}$, being a direct connection to the density of these exoplanets, can yield a first characterization of their internal structure. Moreover, regarding the atmospheric characterization of these systems, we calculated a quantity that describes the relative atmospheric scale height of their atmospheres, $H$. With the measurement of $H$ in km, we can define the extension of the absorbing annulus due to the planet's atmosphere. We made use of the equations from \cite{Southworth2007}, \cite{Southworth2010}, \cite{Winn2010}, \cite{Seager2011}, \cite{Turner2016}, \cite{Alexoudi2018}, in order to compute all these quantities. 

The surface gravitational acceleration is given by Eq. \ref{equ_gp}: 
\begin{equation}
\mathrm{g_p}=\mathrm{\frac{2\pi} {P}} \left( \frac{a}{R_p} \right)^2 \frac{\sqrt{1-e^2}}{\mathrm{sin}\,i} \mathrm{K_{\star}}.
  \label{equ_gp}
\end{equation}

\noindent

The modified equilibrium temperature $\mathrm{T_{eq}}$ is as follows in Eq. \ref{equ_Teq}:

\begin{equation}
\mathrm{T_{eq}}=\mathrm{T_{eff}} \left( \frac{\mathrm{R_{\star}}}{2\,a} \right)^{1/2},
  \label{equ_Teq}
\end{equation}
\noindent
where $\mathrm{T_{eff}}$ is the effective temperature of the host star. 

We estimated the relative atmospheric scale height of the atmospheres of our sample using Eq. \ref{scale_height} \citep{Winn2010}:,

\begin{equation}
H= \mathrm{\frac{k_\beta \ T_{eq}}{\mu_m \ g_p }},
\label{scale_height}
\end{equation}
where $\mathrm{k_ \beta}$ is Boltzmann's constant, $\mathrm{T_{eq}}$ is the equilibrium temperature of the planet, $\mathrm{\mu_m}$ the mean molecular mass and $\mathrm{g_p}$ the local gravitational acceleration. We adopt a mean molecular mass of approximately $2.3$ amu, which is typical for a hot Jupiter exoplanet with a H/He dominated atmosphere (e.g. \cite{Sing2016}). Planets with large atmospheric scale height of many kms are excellent targets for atmospheric characterization through transmission spectroscopy  (e.g. \cite{Burrows2014, Mallonn2015, Sing2016}). 

Moreover, TESS with uninterrupted datasets is expected to improve on the orbital parameters of those systems, hence we can provide a better constrained impact parameter b value for each one of them. The impact parameter is a quantity that shows the projected distance between the planetary center and the stellar center \citep{SeagerMallenOrnelas2003}, and it is given by Eq. \ref{impact}, where $i$ is the inclination of the system and $a/R_s$ is the normalized semi-major axis in units of stellar radii. 
\begin{equation}
\mathrm{b} = \mathrm{cos(\textit{i})} \times  a/R_s. 
\label{impact}
\end{equation}
\noindent

\section{Results}
We used publicly available datasets from TESS with high-quality photometric precision in order to refine the parameters from ten inflated hot gas giant exoplanets. All the derived parameters from this work for each target are displayed in Tables 3-12, along with their $1\sigma$ uncertainty and a reference with their previously obtained parameters. In the following paragraphs, we shortly review each target, including some of their most important features, and present our results with a direct comparison between the individual investigations.

\subsection{WASP-140 b} 
From the exoplanets observed with TESS and analyzed in this work, the one with the largest radius uncertainty is \mbox{WASP-140 b} (see Table~\ref{sample_params}{\ns}). In the discovery paper \citep{Hellier2017}, \mbox{WASP-140 b} belongs, according to previous observations, to a binary system. WASP-140 A is a K0 star, rather active, with an effective temperature of $\mathrm{T_{eff}(K)}$ = $5300 \pm 100$ and $V_{mag} =11.1$, while \mbox{WASP-140 B} is about $2$ magnitudes fainter. The planet of this system, \mbox{WASP-140 b}, has a mass of 2.4 $\mathrm{M_J}$ and an orbital period of P=2.2 days. It orbits around the host star, WASP-140A, on a grazing, at an impact parameter value b = 0.93, and eccentric ($e = 0.047 \pm 0.004$) orbit. \mbox{WASP-140 b} is a massive exoplanet with a short period  of a significant eccentricity. All these characteristics of a hot Jupiter are not met usually around K-type stars, and maybe the studies of those are the key indicators to understand the magnetic activity of the host \citep{PoppenhaegerWolk2014}. Interestingly, WASP-140 A is considered as a magnetically active star, as there are detected some star-spots on some light curves in the discovery paper. The presence of star-spots on the received light curves are able to alter the correct derivation of the transit depth and the correct measurement of the planetary radius \citep{Morris2018, Oshagh2013}.   

We are interested to investigate this system with TESS and focus on the precise determination of its planetary radius. The large uncertainty of this radius measurement might be attributed to the grazing nature of this system and/or the partial transits reported from ground-based observations. The activity of the host cannot be ruled out, either, as a contributor to this radius uncertainty value.  

TESS observed \mbox{WASP-140 b} during the sectors 4, 5 and 31. A total of 28 available light curves were employed in this work to derive the properties of \mbox{WASP-140 b}. Since, WASP-140 system is a known binary with TESS Input Catalog (TIC) identifiers for both stars, the SPOC pipeline estimates a dilution factor and corrects for the amount of the light contamination in the final light curve \citep{Thomson2016}. This dilution metric is named "CROWDSAP" and it is presented on the header of the TESS target pixel files (TPFs). For \mbox{WASP-140 b}, we obtain an average of 0.85 as a crowding estimation, for all the sectors of observation. This value signifies that the contamination of the received flux by nearby sources, is approximately $14\%$. According to \cite{Guerrero2021}, "CROWDSAP" values of less than 0.8 are not trustworthy regarding photometric measurements, therefore for WASP-140 TESS observations, the resulting light curves can be considered as photometrically reliable, since they are not contaminated severely by the companion. 

In Fig. \ref{wasp140b_corner}, we present the TESS folded light curves of \mbox{WASP-140 b} and the best fit transit model at the upper panel. The derived parameters from the emcee approach are shown on the corner plot, at the lower panel. A direct comparison of this work with the previously published parameters from \cite{Hellier2017} is in Table \ref{wasp140b_table}{\ns}. 
We pinpoint that our results do not match the previous investigation within $1\sigma$. We report a later mid-time point of $1.2$ minutes, while the updated orbital parameters yield a more precise impact parameter value. Even though, the orbital parameters, $i$ and $a/R_s$, are not in agreement either with the previous results, the uninterrupted TESS data, yielded a better acquisition of those measurements. The high-cadence datasets provided better constrained and more precise $i$ and $a/Rs$ values, which in turn yielded a more precise b for this system of $0.85$. However, this newly derived determination of b is associated with a better determination of the transit depth, which now is greatly improved. Moreover, we report a refined measurement of a smaller planetary radius for \mbox{WASP-140 b} by approximately $12\%$. We also derived a better precision on the reported asymmetric radius uncertainty of \mbox{WASP-140 b} by approximately $86\%$ and $67\%$. The uncertainty on the planetary radius has been greatly improved by this investigation and places the planet with conviction to a lower inflated radii regime with a smaller estimated temperature. However, the planet remains well beyond the cutoff temperature of 1000K for the inflation to happen \citep{MillerFortney2011}, and it continues to have an excess in its radius compatible with its temperature levels. 

Exoplanets of grazing transits around moderately bright hosts are difficult to parameterize from the ground, as in the case of WASP-168 b in \cite{Hellier2019}. Nevertheless, in this work, using TESS, we improved significantly on the parameters of \mbox{WASP-140 b}, that now it is characterized as a less puffy and more dense exoplanet than it was thought to be.  

\begin{table*}
\begin{center}
\caption{Physical properties of \mbox{WASP-140 b} as derived in this work from the modeling of the TESS light curves and the emcee analysis , in comparison to the previously published work of \cite{Hellier2017}.} 
\label{wasp140b_table}
\begin{tabular}{cccc}
\hline
\noalign{\smallskip}
&&This work &\cite{Hellier2017} \\
\noalign{\smallskip}
& Parameters [units] & Values $\pm 1\sigma$ & Values $\pm 1 \sigma $ \\
\hline
\noalign{\smallskip}
&$\mathrm{T_c ~[BJD_{TDB}]}$ &$2456912.35261 \pm 0.00016$ &$2456912.35183 \text{*} \pm 0.00015$  \\ 
&$i$ [$^{\circ}$] &$84.30 \pm 0.06$ &$83.3\substack{+0.5\\-0.8}$ \\ 
&$\mathrm{a/R_s} $&$8.58 \pm 0.06$& $7.98 \pm 0.39$ \\ 
&$\mathrm{R_p / R_s}$  &$0.1464\pm0.0010$ & $0.1656\text{**}\substack{+0.0494\\-0.0216}$\\ 
& b & $0.851 \pm 0.011$ & $0.93 \substack{+0.07\\-0.03}$ \\ 
& $\mathrm{R_p ~ [R_J]}$&$1.27 \pm 0.06$ &$1.44 \substack{+0.42 \\ -0.18}$\\ 
&$\mathrm{ \rho_p ~[\rho_J]}$ &$1.19\pm 0.17$& $0.8 \pm 0.4$ \\ 
& $\mathrm{\log{g_p} ~[cgs]}$ &$3.592\pm 0.009$ & $3.4\pm 0.2$\\ 
& $\mathrm{T_{eq} ~[K]}$&$1270 \pm 25$  &$1320 \pm 40$ \\ 
& $H$ ~[km] & $123 \pm 4$ & - \\
\hline 
\end{tabular}
\end{center}
 \begin{threeparttable}
\begin{tablenotes}
      
      \item * Converted from Heliocentric Julian Date (HJD) of $2456912.35105\pm 0.00015$ to $\mathrm{BJD_{TDB}}$
      \item ** From the quoted $\mathrm{R_p}$ and $\mathrm{R_s}$ values;  $\mathrm{R_p}$ was poorly constrained since the fitted b makes the transit grazing. 
    \end{tablenotes}
 \end{threeparttable}
\end{table*}

\subsection{WASP-136 b}

Previous studies on \mbox{WASP-136 b}, have shown that this exoplanet belongs to an interesting category of planets, as a short-period hot Jupiter that orbits a sub-giant star. The limited candidates of this population might be attributed to tidal disruption that causes planets to spiral inwards and to the star \citep{Lam2017}. 

\mbox{WASP-136 b} completes a full orbit around its evolved \mbox{late-F} host star ($V_{mag}=9.93$) in 5.22 days and, with a radius of $1.38\,\pm\,0.16\,\mathrm{R_J}$ and mass of $1.51\,\pm\,0.08\,\mathrm{M_J}$, it is an inflated giant planet, which is half as dense as Jupiter. Intriguingly, \mbox{WASP-136} is at its final main sequence phase and the derived planetary radius in the work of \cite{Lam2017} found to be $25\%$ larger than expected in the models of \cite{Fortney2007}. One plausible explanation is that the star moves towards the sub-giant branch and the intensity from the irradiation on the planet is expected to increase dramatically, the planet can heat up, through the stellar irradiation trapped in the interior of the planet, and trigger another re-inflation. A precise radius estimation might provide further clarifications on what the current status of this exoplanet's radius extension is. 

We made use of six full transits of \mbox{WASP-136 b}, from the publicly available TESS 2-minutes cadence data in order to update/confirm the previous findings from the discovery paper \citep{Lam2017}, where one full and two partial transits were analyzed. We did not take into consideration the last transit light curve of sector 42, due to a starspot crossing throughout the transit chord that would lead to a biased measurement of the transit depth. 

The final modeled light curve with the associated residuals are presented at the upper panel in Fig. \ref{wasp136b_corner}, while the best fit parameters are shown in the corner plot at the lower panel. A comparison of the findings in this work with the previous complete study of \cite{Lam2017} is presented in Table \ref{wasp136b_table}{\ns}. The derived $\mathrm{R_p/R_s}$ varies significantly from the previous reported value by more than $3\sigma$, indicating deeper transit light curves and larger planetary radius. However, the derived planetary radius is in agreement with the previously reported value, overall. We find an additional difference of $9\%$ in the radius measurement between the two investigations, that makes it a total of $34\%$ larger than expected in \cite{Fortney2007}. The uncertainty on the planetary radius of this exoplanet has only been improved marginally with the analysis of the TESS data. We conclude to a larger planetary radius that yields a slightly more inflated and less dense exoplanet, that is consistent with the previous work of \cite{Lam2017}.    

\begin{table*}
\begin{center}
\caption{Physical properties of \mbox{WASP-136 b} derived in this work with TESS data, in comparison to the previously published work of \cite{Lam2017}.}
\label{wasp136b_table}
\begin{tabular}{cccc}
\hline
\noalign{\smallskip}
&&This work &\cite{Lam2017} \\
\noalign{\smallskip}
& Parameters [units] & Values $\pm 1\sigma$ & Values $\pm 1\sigma$ \\
\hline
\noalign{\smallskip}
&$\mathrm{T_c ~[BJD_{TDB}]}$  & $2456776.9055 \pm 0.0011$ & 2456776.90615 $\pm 0.0011$  \\ 
&$i$ [$^{\circ}$]     &$87.7 \pm 1.2$               &$84.7\substack{+1.6\\-1.3}$ \\ 
&$\mathrm{a/R_s}  $          &$7.4 \pm 0.3$                & $6.43 \pm 0.65$ \\ 
&$\mathrm{R_p / R_s}$            &$0.0680 \pm 0.0005$          &$0.0641 \pm 0.0012$ \\ 
& b                &$0.30 \pm 0.16$              &$0.59 \substack{+0.08\\-0.14}$ \\ 
& $\mathrm{R_p ~ [R_J]}$     &$1.50 \pm 0.15$              &$1.38 \pm 0.16$\\ 
&$\mathrm{ \rho_p ~[\rho_J]}$  &$0.44\pm 0.14$               &$0.58 \substack{+0.23\\ -0.15}$ \\ 
& $\mathrm{\log{g_p} ~[cgs]}$   &$3.29 \pm 0.05 $             & $3.26\pm 0.09$\\ 
& $\mathrm{T_{eq} ~[K]}$      &$1630 \pm 40 $               &$ 1742\pm 82$ \\ 
& $H$ ~[km]           &$310\pm 40$                  &- \\
\hline 
\end{tabular}
\end{center}
\end{table*} 

\subsection{WASP-113 b}

\mbox{WASP-113 b} is a hot Jupiter that orbits a G1 type host star in a period of about $4.5$ days. In the work of \cite{Barros2016}, it is shown that the planet has a mass of \mbox{$\mathrm{M_p}=0.48\, \mathrm{M_J}$} and an inflated radius of \mbox{$\mathrm{R_p}=1.41\,\mathrm{R_J}$}, hence a density of about \mbox{$\mathrm{\rho_p} = 0.172\,\mathrm{\rho_J}$}. However, the planetary radius was expected about $2 \sigma $ smaller in the work of \cite{Fortney2007}, assuming a coreless model. In our work, using TESS, we can confirm/refine the radius measurement of \mbox{WASP-113 b}, and investigate if we can improve on its radius uncertainty.    

For this purpose, we analyzed a total of eight transit events (one partial transit was rejected), as they were observed with TESS during sectors $23$ and $24$. The best-fit model is shown on the upper panel in Fig. \ref{wasp113b_corner}, while the best fit parameters are depicted on the corner plot at the lower panel, along with the values of the detrending coefficients. In Table \ref{wasp113b_table}{\ns}, we present the comparison of the newly derived parameters with the previous work of \cite{Barros2016}. The radius measurement is in agreement between the two independent investigations, along with the rest of the parameters that were confirmed now with independent datasets provided by TESS. Even though, the TESS data analysis improved only marginally the orbital parameters of this system, however it confirmed an exoplanet of a largely extended atmosphere ($H$>1000 Km) that orbits a bright host star, i.e., an excellent target for transmission spectroscopy investigations. This is one of the cases where ground-based observations and space-based ones, came into a complete agreement, highlighting this way the good function of the ground- and space-synergy. 

\begin{table*}
\begin{center}
\caption{Physical properties of \mbox{WASP-113 b} derived in this work from TESS data, in comparison to the previously published work of \cite{Barros2016}.}
\label{wasp113b_table}
\begin{tabular}{cccc}
\hline
\noalign{\smallskip}
&&This work &\cite{Barros2016} \\
\noalign{\smallskip}
&Parameters [units] &Values $\pm 1\sigma$ &Values $\pm 1\sigma$ \\
\hline
\noalign{\smallskip}
&$\mathrm{T_c ~[BJD_{TDB}]}$ &$2457197.09751 \pm 0.00045$ 
& $2457197.098226 \text{*} \pm 0.000040$  \\ 
&$i$ [$^{\circ}$]    &$86.9 \substack{ +1.4\\-1.0}$        &$86.46 \substack{+1.2\\-0.64}$ \\ 
&$\mathrm{a/R_s}$          &$8.2 \substack{ +0.6\\-0.5}$            &$7.87\pm0.59$ \\ 
&$\mathrm{R_p / R_s}$           &$0.0917 \substack{ +0.0014\\-0.0013}$       &$0.0899 \pm 0.0015$ \\ 
&b               &$0.45 \pm0.20$            &$0.486 \substack{+0.063\\-0.14}$\\
& $\mathrm{R_p ~ [R_J]}$    & $1.47 \pm 0.11 $         &$1.409 \substack{+0.095\\-0.14}$\\ 
&$\mathrm{ \rho_p ~[\rho_J]}$ &$0.15 \pm0.04$            & $0.172 \substack{+0.055\\-0.034}$\\
& $\mathrm{\log{g_p} ~[cgs]}$&$2.73 \pm0.08$            &$2.744 \substack{+0.081\\-0.072}$\\
& $\mathrm{T_{eq} ~[K]}$        &$ 1460\pm 70 $            &$1496 \pm 60$ \\ 
& $H$ ~[km]           &$1030 \pm 230$            & - \\
\hline 
\end{tabular}
\end{center}
\begin{threeparttable}
\begin{tablenotes}
      
      \item * Converted from HJD of $2457197.097459\pm 0.00004$ to $\mathrm{BJD_{TDB}}$
    \end{tablenotes}
 \end{threeparttable}
\end{table*}

\subsection{WASP-120 b}

\mbox{WASP-120 b} was discovered and characterized with the work of \cite{Turner2016}. Five ground-based transit observations were used to determine the system's parameters, which yielded a massive exoplanet of $\mathrm{M_p}=4.85\,\mathrm{M_J}$ and an inflated radius of $\mathrm{R_p}=1.73\,\mathrm{R_J}$. The eccentricity of \mbox{WASP-120 b} is significant and equal to $e=0.059 \pm 0.02$. Moreover, this exoplanet has an orbital period of $3.6$ days, around its F5 type host. The host star is bright with $V_{mag}= 11$, an age of $0.7$ Gyr and rather important activity \citep{Turner2016}. This activity was pinpointed due to an observed difference, of $1.2 \pm 0.4 \times 10^{-3}$, between the transit depths of two light curves. Another explanation for this difference could be the presence of a nearby companion. Interestingly, in the work of \cite{Bohn2020}, there are hints that the host star belongs to a hierarchical triple system. 

TESS revisited this system with four observing sectors. In our work, we employed 24 transit light curves of \mbox{WASP-120 b}, in total. We analyzed the TESS data as for the previous targets of our sample, and we kept the significant eccentricity of this planet as a fixed parameter during the transit light curve fit process. Our results have shown a smaller planetary radius of $\mathrm{R_p} = 1.39\,\pm\, 0.08\,\mathrm{ R_J}$ for \mbox{WASP-120 b}, in agreement with the previously published value at $1 \sigma$. Also, we report an earlier mid-time transit point by $2.3$ minutes, and considering that \mbox{WASP-120 b} is the heaviest exoplanet of our sample, this might be an evidence that transit timing variations (TTVs) effects take place in this system. The large mass of the planet and the low metallicity of the host star indicate that the planet received a large portion of radiation from its host in order to puff-up and extend its radius. However, WASP-120 was reported to obtain significant activity in \cite{Turner2016} or, possibly, a companion star. The reported difference between two transit depths of independent observations is confirmed in our work with the analysis of the TESS datasets. More precisely, we fit the 24 light curves individually, with the same emcee process which is based on a MCMC approach, as for the rest of our targets, and obtained the fit transit depths for each of the 24 transit light curves of \mbox{WASP-120 b}. The largest transit depth difference occurs between the transits at the mid-time points of \mbox{$2458426.172325$ $\mathrm{BJD_{TDB}}$} and \mbox{$2458447.839933$ $\mathrm{BJD_{TDB}}$}, at the observations of sector $4$ and sector $5$, respectively, and it is approximately $1.1 \pm 0.5 \times 10^{-3}$. We demonstrated that this difference between transit events is repeated and can be attributed to the stellar activity, in agreement with the previous investigation by \cite{Turner2016}. Moreover, we can exclude significant contamination of the datasets due to a companion star by examining the crowding metric, "CROWDSAP", for all the observing sectors. This metric indicates that on average, for all the four sectors of this observation, only $4\%$ of the received light has been obtained from nearby sources. By inspection of the TPF for this exoplanet from TESS, we observed that the companion falls in the detector's field-of-view, hence the SPOC team took this third light contamination into account in the derivation of the PDCSAP flux. This amount of contamination of $4\%$ can be considered in practice negligible as the planet-to-star radius ratio would be underestimated only by $0.02\%$, hence the aforementioned transit depth difference might be attributed to the host activity and not to the presence of the nearby companion.  

Furthermore, with the analysis of the TESS data, we improved the precision of the orbital parameters of this system significantly. The $i$ and $a/R_s$ are not in agreement with the previous work of \cite{Turner2016}, for more than $2\sigma$. Our analysis yielded an updated impact parameter for this system that differs from the previous one for almost $3\sigma$. The larger semi major axis, as derived from the fit, supports the presence of a smaller equilibrium temperature, that deviates from the literature value by approximately $3\sigma$, which consequently yields a reduced atmospheric scale height equal to half its initial value. In Fig. \ref{wasp120b_corner}, the final model fit is presented. The parameterization of the system after the emcee process and fit is shown on a corner plot at the lower part of the same figure, while a comparison with the work of \cite{Turner2016}, is presented in Table \ref{wasp120b_table}{\ns}. We conclude to a slightly smaller, denser and with a thin atmosphere of less than \mbox{$H=100\,\mathrm{km}$} exoplanet. 

\begin{table*}
\begin{center}
\caption{Physical properties of \mbox{WASP-120 b} derived in this work with TESS data, in comparison to the previously published work of \cite{Turner2016}.}
\label{wasp120b_table}
\begin{tabular}{cccc}
\hline
\noalign{\smallskip}
&&This work &\cite{Turner2016} \\
\noalign{\smallskip}
&Parameters[units] &Values $\pm 1\sigma$ &Values $ \pm 1\sigma$ \\
\hline
\noalign{\smallskip}
&$\mathrm{T_c ~[BJD_{TDB}]}$ &$2456779.4347 \pm 0.0005$ & $   2456779.4363 \text{*}\pm0.0005$  \\ 
&$i$ [$^{\circ}$] &$84.54 \pm 0.35$ &$82.54 \pm 0.78$ \\ 
&$\mathrm{a/R_s}$ &$6.8 \pm 0.2$& $5.90 \pm0.339$ \\ 
&$\mathrm{R_p / R_s}$  &$0.0751 \pm 0.0005$ &$0.08093 \pm 0.00099$\text{**} \\ 
& b &$0.65 \pm 0.05$ &$0.78 \pm 0.02$   \\ 
& $\mathrm{R_p ~ [R_J]}$&$1.39 \pm 0.08$ &$1.473\pm 0.096$\\ 
&$\mathrm{ \rho_p ~[\rho_J]}$&$1.76\pm 0.32$& $ 1.51 \substack {+0.33\\ -0.26}$ \\ 
& $\mathrm{\log{g_p} ~[cgs]}$ &$3.86 \pm 0.03$ & $ 3.707 \pm 0.056$\\ 
& $\mathrm{T_{eq} ~[K]}$&$1749 \pm 41 $& $ 1880 \pm 70$ \\ 
& $H$ ~[km]&$90\pm 7$ &- \\
\hline 
\end{tabular}
\end{center}
\begin{threeparttable}
\begin{tablenotes}
      
      \item * Converted from HJD of $2456779.43556\pm 0.00051$ to $\mathrm{BJD_{TDB}}$
    \item ** Computed from the given $\mathrm{(R_p/R_s)}^2 = 0.00655\pm0.00016$
       
    \end{tablenotes}
 \end{threeparttable}
\end{table*} 

\subsection{WASP-93 b}
\mbox{WASP-93 b} was discovered and characterized in the work of \cite{Hay2016}. \mbox{WASP-93 b} has a mass of \mbox{$\mathrm{M_p} = 1.47\, \mathrm{M_J}$} and it orbits a F4 star with period of about 2.73 days. Later on, in \cite{Gajdos2019}, the estimated radius of \mbox{WASP-93 b} was much smaller than the value published by \cite{Hay2016}, $\mathrm{R_p} = 0.0873 \pm 0.0025\,\mathrm{R_s}$  and $\mathrm{R_p} = 0.1080 \pm 0.0059\, \mathrm{R_s}$, respectively. This inconsistency of more than $3\sigma $ regarding the planetary radii gave us additional motivation to revisit \mbox{WASP-93} system and re-evaluate this transit depth measurement with TESS. Therefore, we made use of TESS light curves to compare the findings with this previous estimation, one of which is based on only one transit event \citep{Gajdos2019}. We used TESS observations of sector 17 in order to revisit and refine the parameters of \mbox{WASP-93 b}. Grazing systems are hard to parameterize due to the fact that their transit light curves have rounded bottoms, constraining this way the information on the orbital parameters derived from a well-defined ingress and egress (when the planet starts to cross the projection of the stellar disk and when it exits). The planetary radii are sometimes poorly defined in grazing systems, but we expect TESS to provide insights on this domain. 

For \mbox{WASP-93 b}, in total, there were employed six light curves, cleaned from outliers and detrended from systematics. We analyzed them similarly to our other targets of this investigation. In Fig. \ref{wasp93b_corner}, we present the best-fit model on the folded light curves on the upper panel. In \cite{Hay2016}, it is pinpointed that the mid-times are not well-defined due to uncertainties in the transit ephemeris. Hence, we used in our analysis the revised ephemeris in \cite{Gajdos2019} ($\mathrm{T_c}=2456079.553552\pm 0.00457\,\mathrm{BJD_{TDB}} $). We observed an offset in our transits' ingress of approximately $9.3$ minutes later. The almost v-shaped transit gives indications of a background star that might contribute significantly to the received light curve. However, there are no companions with important brightness in the proximity in order to produce such effect, while another scenario could be that \mbox{WASP-93 b} belongs to a triple system \citep{Hay2016}. The final parameters of the system are shown at the corner plot in Fig. \ref{wasp93b_corner}, while the comparison with the works of \cite{Hay2016} and \cite{Gajdos2019} are presented in Table \ref{wasp93b_table}{\ns}. The planetary radius of $\mathrm{R_p} = 1.54 \pm 0.06\, \mathrm{R_J}$  derived in our work is consistent with the work of \cite{Hay2016} of $\mathrm{R_p}=1.597 \pm 0.077\, \mathrm{R_J}$, within $1 \sigma$. The derived parameters agree broadly with the published ones, while the transit depth and planetary radius are significantly different from in the work of \cite{Gajdos2019} with $\mathrm{R_p/R_s}= 0.0873\pm0.0025$ and $\mathrm{R_p}=1.29 \pm 0.05\, \mathrm{R_J}$. This TESS investigation improved only marginally on the planetary radius uncertainty, it confirmed the findings in \cite{Hay2016} and solved a discrepancy between the two reported transit depths for \mbox{WASP-93 b}. Interestingly, the orbital parameters are consistent within $1\sigma$, supporting a grazing exoplanet in all of these works, highlighting the need to parameterize grazing systems with caution. In a recent work published by \cite{Wong2021}, the planetary radius and orbital parameters are in complete agreement with our work. 

\begin{table*}  
\begin{center}
\caption{Physical properties of \mbox{WASP-93 b} derived in this work with TESS data, in comparison to the previously published works of \cite{Hay2016} and \cite{Gajdos2019}.}
\label{wasp93b_table}
\begin{tabular}{ccccc}
\hline
\noalign{\smallskip}
&&This work &\cite{Hay2016} &\cite{Gajdos2019} \\
\noalign{\smallskip}
&Parameters [units] &Values $\pm 1\sigma$ &Values $\pm 1\sigma$ &Values $\pm 1\sigma$\\
\hline
\noalign{\smallskip}
&$\mathrm{T_c ~[BJD_{TDB}]}$ &$2456079.560 \pm 0.0046$ & $   2456079.5650 \text{*} \pm 0.0004$& $2456079.553552 \text{*}\pm 0.00457$ \\ 
&$i$ [$^{\circ}$] &$81.55 \pm 0.32 $& $81.18 \pm0.29$ &$82.27 \pm 0.58$\\ 
&$\mathrm{a/R_s}$&$6.11\substack{+0.18 \\-0.15}$& $5.94\pm0.13$ &$6.45\pm0.35$\\ 
&$\mathrm{R_p / R_s}$   &$0.1017\substack{+0.0028 \\-0.0020}$ & $0.10474\pm0.00062 \text{**}$ &$0.0873\pm0.0025$\\ 
& b & $0.90 \pm 0.04$ & $ 0.904 \pm0.009$  &$0.868\pm0.080$\\ 
& $\mathrm{R_p ~ [R_J]}$&$1.54 \pm 0.06$ &$ 1.597\pm 0.077$ &$1.29\pm 0.05\text{***}$\\ 
&$\mathrm{ \rho_p ~[\rho_J]}$&$0.40\pm 0.09$& $  0.360 \pm 0.084$&- \\ 
& $\mathrm{\log{g_p} ~[cgs]}$&$3.17 \pm 0.09$ & $  3.120 \pm0.093$ &-\\ 
& $\mathrm{T_{eq} ~[K]}$ &$1910 \pm 40 $& $1942 \pm 38$ &-\\ 
& $H$ ~[km] & $480 \pm 110$ & - &-  \\
\hline 
\end{tabular}
\end{center}
 \begin{threeparttable}
\begin{tablenotes}
      
      \item * Converted to $\mathrm{BJD_{TDB}}$ from $2456079.56420 \pm 0.00045$ and $2456079.55280\pm0.00457$ HJD \citep{Gajdos2019}. 
      \item ** Computed from the given $\mathrm{(R_p/R_s)}^2 = 0.01097 \pm 0.00013$
      \item ***From the quoted $\mathrm{R_p}$ estimation in units of Earth radii provided in that work.  
    \end{tablenotes}
 \end{threeparttable}
\end{table*}

\subsection{HAT-P-16 b}
\mbox{HAT-P-16 b} is a transiting inflated hot Jupiter exoplanet that orbits a bright F8 dwarf ($V_{mag}= 10.8$) every 2.77 days on an eccentric orbit ($e=0.036$). This is the second heaviest exoplanet of our sample. From previous investigations on this exoplanet \citep{Buchhave2010}, the planetary radius is estimated as $\mathrm{R_p} = 1.289 \pm 0.066\,\mathrm{ R_J}$, its planetary mass as $\mathrm{M_p} = 4.193 \pm 0.094\,\mathrm{ M_J}$ and its density $\mathrm{\rho_p} = 2.42 \pm 0.35\, \mathrm{g\, cm^{-3}}$. 

Using seven light curves obtained with TESS, we revisited this system and defined its properties. We kept the eccentricity (as with the previously highly eccentric systems) fixed to the literature value during the transit light curve fit process. The best fit transit model is shown at the top panel of Fig. \ref{hatp16b_corner}, and the associated best fit parameter corner plot on the lower panel of that same figure. At Table \ref{tess_observations_hatp16b}{\ns}, we present the comparison of our work with the investigations by \cite{Buchhave2010} and \cite{Ciceri2013}.  With the analysis of the TESS datasets, we report a later transit by 15 minutes for \mbox{HAT-P-16 b}, an indication that probable TTVs are present. Since it is a massive exoplanet, TTVs are not unusual in this case. The derived orbital parameters are in a broad agreement with the works of \cite{Ciceri2013} and \cite{Buchhave2010}, while the impact parameter is consistent within $1\sigma$ with the work of \cite{Ciceri2013}. The newly derived planetary radius is in better agreement with the work of \cite{Buchhave2010}, within $1\sigma$. Even though, with our TESS investigation we did not improve on the planetary radius uncertainty of this system, however we confirmed a planet with the same density and surface gravity as thought in \cite{Buchhave2010}. In the future, TESS is not expected to observe \mbox{HAT-P-16 b} again.

\begin{table*}
\begin{center}
\caption{Parameter refinement of \mbox{HAT-P-16 b} using high-cadence TESS data. The $\mathrm{R_p}$ value as derived from the analysis in our work is in better agreement with \cite{Buchhave2010} within $1\sigma $ significance.} 
\label{tess_observations_hatp16b}
\begin{tabular}{ccccc}
\hline
\noalign{\smallskip}
\centering
%\noalign
{\smallskip}
   & & This work  &\cite{Buchhave2010} &\cite{Ciceri2013} \\
\noalign{\smallskip}
&Parameters [units] & Values $\pm 1\sigma$ &Values $\pm 1\sigma$ &Values $\pm 1\sigma$  \\
\hline
\noalign{\smallskip}
&$\mathrm{T_c ~[BJD_{TDB}]}$ & 2455027.60356 $\pm0.00035 $ & $2455027.59293 \pm 0.00031$ & $2455027.59281 \pm 0.00040$ \\ 
&$i$ [$^{\circ}$] &$88.4	\pm 1.0$&$86.6 \pm0.7$ &$87.74 \pm 0.59$ \\ 
&$\mathrm{a/R_s}$ &$7.73\pm	0.23$&$7.17 \pm0.28$ &$7.67 \pm 0.18\text{*}$\\ 
&$\mathrm{R_p / R_s}$  &$0.1063\pm	0.0007$&$0.1071\pm0.0014$ &$0.1067 \pm 0.0014$\\ 
&b &$0.220\pm	0.130$&$0.439 \substack{+0.065 \\-0.098}$ & $0.30 \pm 0.08\text{*}$ \\ 
& $\mathrm{R_p ~ [R_J]}$ &$1.31\pm	0.06$&$1.289 \pm0.066$ &$ 1.190 \pm 0.035$ \\ 
&$\mathrm{ \rho_p ~[\rho_J]}$ &$1.87\pm0.25  $ &$ 1.95 \pm0.28 \text{**}$&$2.33 \pm0.20$\\
& $\mathrm{\log{g_p} ~[cgs]}$&$3.804\pm0.027$&$3.8 \pm0.04$& $3.87 \pm 0.024$\\ 
& $\mathrm{T_{eq} ~[K]}$ &$1566	\pm31$ &$1626 \pm 40$&$1567 \pm 22$\\
& $H$ ~[km] &$93 \pm 7$ &- &-  \\
\hline 
\end{tabular}
\end{center}
 \begin{threeparttable}
\begin{tablenotes}
      
      \item * Estimated from the given values for the angular separation of the system in AU and for the inclination in \cite{Ciceri2013}.
      \item ** Converted to units of Jupiter's density from $\mathrm{\rho_p}= 2.42\pm 0.35\,\mathrm{g\,cm^{-3}}$ \citep{Buchhave2010}. 
    \end{tablenotes}
 \end{threeparttable}
\end{table*} 

\subsection{WASP-123 b}
\mbox{WASP-123 b} orbits a bright ($V_{mag} = 11.1$) G5 star, every 3 days. It is a hot Jupiter with an inflated radius of $\mathrm{R_p}= 1.32\,\mathrm{ R_J}$. Considering the measurement of its mass of $\mathrm{M_p}=0.9\,\mathrm{M_J}$ , the planet has a density of about $ \mathrm{\rho_p} = 0.4\,\mathrm{ \rho_J}$ \citep{Turner2016}. It is a rather typical hot Jupiter and the revision of its planetary radius will contribute to the understanding of the general picture of this kind of exoplanets around evolved stars.   

The analysis of the TESS datasets based on observations on this exoplanet from the sectors 13 and 27, interestingly, showed that there is a discrepancy regarding the derived transit depths. In sector 13, we derived an $\mathrm{R_p/R_s}=0.1240 \pm 0.0012$, while in sector 27, $\mathrm{R_p/R_s}=0.1054 \pm 0.0008$. To investigate this case, we made use of the "CROWDSAP" metric, that defines the ratio between the target flux and the total flux that falls on the photometric aperture. The PDCSAP fluxes are usually adjusted according to this crowding factor. For \mbox{WASP-123 b} observations at sector 13, the CROWDSAP metric is about $69.3\%$, while for sector 27, it is $ 89.8\% $ and this indicates, for sector 13, a contamination due to third light of approximately $30\%$, while for sector 27, of about $10\%$. In \cite{Guerrero2021}, it is pinpointed that "CROWDSAP" values of less than 0.8 yield an unreliable photometry. However, this metric can be sometimes underestimated by the PDC pipeline and can yield an overestimated third light contamination, as in the work of \cite{Parviainen2021} for the exoplanet \mbox{TOI-519 b}. Sector 27 observations support an exoplanet much less contaminated from third sources than sector 13. Based on this, we expect the transits in sector 13 to yield a biased parameter refinement. 

We checked the TOI-catalog \footnote{https://tev.mit.edu/data/}, to investigate further this discrepancy and find the reason of its occurrence. As a public comment, it is mentioned that the transits of \mbox{WASP-123 b} affected the star with TIC31858844, for the entire sector observation. In the work of \cite{Bohn2020}, the companion to \mbox{WASP-123} has only a separation of $4.8''$. It has a stellar mass of $0.4 \mathrm{M_s}$ and effective temperature $\mathrm{T_{eff}(K)}$= 3524. One scenario is that this is highly active M2 dwarf that is blended in the photometry of its companion during the transiting event of \mbox{WASP-123 b}, hence the lower flux that was received in the photometric aperture from \mbox{WASP-123}. We neglected the transits of sector 13, since we cannot account for the missing light, and in our analysis we used only the data from sector 27, which contains transit observations with better target focused photometry. The TIC number of the stellar companion is registered in the TOI-catalog, indicating that the crowding factor is taken into account for the correction of the final PDCSAP light curve. 

In total, we employed seven light curves for our analysis. The best model fit is shown at the upper left panel in Fig. \ref{wasp123b_corner}  (while the best fit model from sector 13 investigation is presented on the right panel for comparison), and the final parameters are presented at the corner plot at the lower panel of the same figure. The comparison with the previous work of \cite{Turner2016} is shown in Table \ref{wasp123b_table}{\ns}. Our TESS findings, other than a later transit midpoint by $2.3$ minutes, are completely consistent with this previous work, even though they did not show any significant improvement on the uncertainty on the planetary radius of this exoplanet either. 

\begin{table*}
\begin{center}
\caption{Physical properties of \mbox{WASP-123 b} derived in this work with TESS data solely from Sector 27 (see in text), in comparison to the previously published work of \cite{Turner2016}.}
\label{wasp123b_table}
\begin{tabular}{cccc}
\hline
\noalign{\smallskip}
&&This work &\cite{Turner2016} \\
\noalign{\smallskip}
& Parameters[units] & Values $\pm 1\sigma$ &Values $\pm 1\sigma$ \\
\hline
\noalign{\smallskip}
&$\mathrm{T_c ~[BJD_{TDB}]}$ &$2456845.1735 \pm 0.0004$ & $2456845.171610  \pm 0.00039$ \text{*}  \\ 
&$i$ [$^{\circ}$] &$86.2 \pm 0.5 $ &$  85.74\pm 0.55$ \\ 
&$\mathrm{a/R_s}$ &$7.29 \pm 0.20$& $7.13 \pm0.25$ \\ 
&$\mathrm{R_p / R_s}$  &$0.1053 \pm 0.0007$ &$0.10536\pm0.00128$ \text{**} \\ 
& b &$0.48 \pm 0.06$ &$  0.530 \pm 0.049 $ \\ 
& $\mathrm{R_p ~ [R_J]}$&$1.35 \pm 0.05$ &$ 1.318 \pm 0.065$\\ 
&$\mathrm{ \rho_p ~[\rho_J]}$&$0.37\pm 0.05$& $   0.393 \pm 0.056$ \\ 
& $\mathrm{\log{g_p} ~[cgs]}$ &$3.063 \pm 0.026 $ & $  3.07 \pm 0.04$\\ 
& $\mathrm{T_{eq} ~[K]}$&$1500 \pm 40 $&$1520 \pm 50$ \\ 
& $H$ ~[km] &$490 \pm 40 $ &$ -$ \\
\hline 
\end{tabular}
\end{center}
 \begin{threeparttable}
\begin{tablenotes}
      \item * Converted in $\mathrm{BJD_{TDB}}$ from $2456845.17082 \pm 0.00039$ HJD in \cite{Turner2016}.
      \item ** Computed from the given $\mathrm{(R_p/R_s)}^2 = 0.01110\pm0.00027$ in \cite{Turner2016}.

    \end{tablenotes}
 \end{threeparttable}
\end{table*}

\subsection{WASP-76 b}

\mbox{WASP-76 b} was discovered and characterized with the work of \cite{West2016}. This exoplanet is an inflated hot Jupiter ($\mathrm{R_p} = 1.83\,\mathrm{ R_J}$) that orbits its bright ($V_{mag}=9.5$) F7 host in 1.8 days. In the recent work of \cite{Southworth2020}, the authors wanted to account for the partial transits of the discovery paper that they were highly contaminated with red noise, and they proceeded with a refinement of this system using one full transit light curve of \mbox{WASP-76 b} obtained from CAHA 1.23m. The authors have discovered that the transit occurred $8.6$ minutes earlier than predicted and the orbital period was $0.54$s shorter than the expected value. Hence, it is suggested that TTVs might be responsible for this deviation. Moreover, they reported a larger radius ($\mathrm{R_p/R_J}=1.885 \substack{+0.117\\-0.042}$) and a higher equilibrium temperature ($\mathrm{T_{eq}} = 2235 \substack{+56\\-25}$) in comparison to the discovery paper. 

During the previous years, numerous studies were conducted on this exoplanet, interestingly some of them using large ground-based telescopes such as the Very Large Telescope (VLT)  \citep{Ehrenreich2020} and/or space telescopes such as the Hubble Space Telescope (HST) \citep{vonEssen2020}. Observations with TESS are currently available to contribute to the understanding of the \mbox{WASP-76 b} planetary system. For this purpose, we made use of TESS observations during sectors 30 and 42. In total, 23 light curves were employed in our analysis. Interestingly, we derived a new transit mid-time, as a free parameter from our MCMC analysis, using emcee, and it appears to occur approximately $14$ minutes earlier than expected from \cite{West2016}, while the orbital period is found to agree with \cite{Southworth2020} and be $0.47$s shorter. We, therefore, confirm the need for a TTV analysis for this system. Moreover, even though our analysis did not improve on the planetary radius uncertainty of this exoplanet either, the high photometric quality of TESS yielded a planetary radius of $1.83\pm 0.04\,\mathrm{R_J}$, which is in complete agreement with \cite{West2016}. However, the analysis in order to derive the transit depth in \cite{West2016} neglects completely the contribution of \mbox{ WASP-76A}'s companion. \mbox{WASP-76B} has a $0.4438''$ separation, and it is $2.58$ times fainter than \mbox{WASP-76A}. The flux contrast between the two stars is approximately 10 \citep{Ehrenreich2020}. Using Eq. 5 in \cite{Ciardi2015}, we applied a correction factor of $\sqrt{1.1}$ to our transit depth, which is compatible with the value of 1.4 that it is suggested in \cite{Ciardi2015}, as a typical factor to apply, to account for the contamination of the third light when the planet transits the primary (instead of the secondary) star. This correction yielded a new $\mathrm{R_p/R_s} = 0.111425 \pm 0.000094$, and moreover a new planetary radius of $\mathrm{R_p }= 1.92 \pm 0.04$, which is in complete agreement with the latest works on this subject \citep{Ehrenreich2020, Southworth2020, vonEssen2020}. The corrected planetary radius yields also a new density estimation of $\mathrm{\rho_{p}} =0.13\pm0.01$ (consequently more parameters are affected by this change in radius and density, and yield a $\mathrm{\log{g_p}} = 2.750 \pm 0.007$, and a $H=1471\pm34$), which is in agreement with the previous within $2 \sigma$, and suggests now a less dense exoplanet. This TESS analysis confirmed the latest reports on this exoplanet, however it also provided a valuable precision in $\mathrm{R_p/R_s}$, in comparison to all the other works, which is very useful for transmission spectroscopy investigations. 
Regarding the TESS light curve of this target, and the crowding metric being "CROWDSAP" = $0.99972$, this signifies that only the $0.3\%$ of this PDCSAP flux comes from contaminating sources. In this case, the analysis of the TESS light curves support the work of \cite{Ehrenreich2020} where they did not confirm the companion either, or, as they suggest, it is completely hiding behind \mbox{WASP-76A} and does not contribute in the received photometry significantly. It is worthy noting that \mbox{WASP-76B} is not registered with an allocated TIC number, which might be a reason why the SPOC team did not consider a dilution factor for this system. 
In Fig. \ref{wasp76b_corner}, we present at the upper panel, the final fit of the TESS light curves along with the associated residuals, while at the lower panel, we present the derived values from the emcee process. We pinpoint here that our work is in a direct comparison to \cite{West2016}, where the authors did not account for the third light contribution, since at the TESS datasets, the companion is not taken under consideration either, or its contribution is not significant, as it appears from the value of the crowding metric. 

\begin{table*}
\begin{center}
\caption{Physical properties of \mbox{WASP-76 b}, as derived in this work with TESS data, in comparison to the previously published work of \cite{West2016}. In brackets, we present the derived values considering the third light contamination.}.
\label{wasp76b_table}
\begin{tabular}{cccc}
\hline
\noalign{\smallskip}
&&This work - [L3 correction] &\cite{West2016} \\
\noalign{\smallskip}
& Parameters[units] &Values $\pm 1\sigma$ &Values $\pm 1\sigma$ \\
\hline
\noalign{\smallskip}
&$\mathrm{T_c ~[BJD_{TDB}]}$ &$2456107.84623 \pm 0.00030$ & $2456107.855819 \text{*}\pm 0.00034$  \\ 
&$i$ [$^{\circ}$] &$89.4 \substack{ +0.5\\-0.6} $&$  88.0 \substack{ +1.3\\-1.6}$ \\ 
&$\mathrm{a/R_s} $&$4.1 \pm 0.01 $& $4.102 \pm0.062$ \\ 
&$\mathrm{R_p / R_s}$  &$0.10624 \pm 0.00009 - [0.111425 \pm 0.000094]$ &$0.10630 \text{**}\pm0.0035$ \\ 
& b &$0.04 \pm 0.05$ &$  0.14\substack{ +0.11\\-0.09} $ \\ 
& $\mathrm{R_p ~ [R_J]}$&$1.83 \pm 0.04 - [1.92 \pm 0.04]$ &$  1.83 \substack{+0.06\\-0.04}$\\ 
&$\mathrm{ \rho_p ~[\rho_J]}$&$0.15\pm 0.01 - [0.13\pm0.01]$& $  0.151 \pm 0.010$ \\ 
& $\mathrm{\log{g_p} ~[cgs]}$ &$2.791 \pm 0.007 - [2.750 \pm0.007]$ & $ 2.80 \pm 0.02$\\ 
& $\mathrm{T_{eq} ~[K]}$ &$2182 \pm 35$ &$2160 \pm 40$ \\ 
& $H$ ~[km] &$1335 \pm 31-[1471\pm34]$ &-\\
\hline 
\end{tabular}
\end{center}
\begin{threeparttable}
\begin{tablenotes}
      
      \item * Converted from $2456107.85507\pm0.00034$ HJD.
      \item **Estimated from the quoted $\mathrm{R_p}$ and $\mathrm{R_s}$ values in \cite{West2016}. 
    \end{tablenotes}
 \end{threeparttable}
\end{table*} 

\subsection{WASP-20 b}
\mbox{WASP-20 b} was discovered and characterized in the work of \cite{Anderson2015}. The binarity of this system (companion at $0.26''$) was analyzed in the work of \cite{Evans2016}, where three scenarios were considered. \mbox{WASP-20 b} is an inflated, Saturn-mass planet. If we ignore the binarity of the system, the planetary radius and planetary mass are \mbox{$\mathrm{R_p} = 1.20 \pm 0.14 \,\mathrm{R_J}$} and \mbox{$\mathrm{M_p}=0.291 \pm 0.017\,\mathrm{ M_J}$}, if the planet transits the brighter star of this system then \mbox{$\mathrm{R_p} = 1.28 \pm 0.15\,\mathrm{ R_J}$} and \mbox{$\mathrm{M_p} = 0.378 \pm 0.022\, \mathrm{M_J}$} and if the planet orbits the fainter star then \mbox{$\mathrm{R_p} = 1.69 \pm 0.12\,\mathrm{ R_J}$} and \mbox{$\mathrm{M_p} = 1.30 \pm 0.19\,\mathrm{ M_J}$}. With TESS, we revisited this interesting system. Using two sectors of TESS observations and a total of ten light curves, we refined the planetary radius to \mbox{$\mathrm{R_p}=1.38 \pm 0.04\,\mathrm{ R_J}$}. Recently, \cite{Southworth2020} tried to determine the planetary radius of \mbox{WASP-20 b} with TESS observations of sector 2. They concluded to \mbox{ $\mathrm{R_p} = 1.382 \pm 0.057\,\mathrm{ R_J}$} for a transit around the brighter star and \mbox{$\mathrm{R_p}=1.69 \pm 0.11\,\mathrm{ R_J}$} for a transit around the fainter star. We used TESS observations of all the available sectors, sector 2 and sector 29, to eventually confirm those findings. The best fit model is shown in Fig. \ref{wasp20b_corner}, at the upper panel therein, and the final parameterization appears on the corner plot at the lower panel. The comparison with the previous work that yielded a full parameterization of the system  \citep{Anderson2015} is presented in the Table \ref{wasp20b_table}{\ns}. 

In TESS observations the companion star is not identified with a TIC number, hence the "CROWDSAP" metric of, an average from both sectors, of 0.997, might not take into account the third light contribution from \mbox{WASP-20B}. Therefore, to account for this contamination of $3\%$, we used the correction factor of $1.34$, as in \cite{Southworth2020} for the analysis of the TESS data of sector 2. We adopted the stellar radius of the work of \cite{Southworth2020} of \mbox{$\mathrm{R_s}=1.242 \pm 0.044\, \mathrm{R_\odot}$ }and with a derived \mbox{ $\mathrm{R_p/R_s} = 0.1153 \pm0.0005$}, which agrees within $1\sigma$ with the previous work on TESS data, we concluded to a planetary radius of $\mathrm{R_p}=1.43 \pm 0.05\,\mathrm{ R_J}$. This planetary radius is in good agreement with both previous works of \cite{Anderson2015} and \cite{Southworth2020}. However, the stellar radii considered in those works are much different with \mbox{ $\mathrm{R_s}=1.392\pm0.044\,\mathrm{ R_\odot}$} and \mbox{$\mathrm{R_s}=1.242 \pm 0.044\,\mathrm{ R_\odot}$} respectively. Taken into account the stellar radius derived in the ground based investigation instead, we concluded to a planetary radius of 1.6 $\mathrm{R_J}$. This is another example of the necessity of the correct parameterization of the stellar parameters in order to obtain accurate planetary parameters. 
For this same system, furthermore, we report a later transit, by approximately $13.7$ minutes, hence TTVs might be present, even though it is unlikely considering the small mass of the planet. The derived planetary radius is in agreement with \cite{Anderson2015}, and also consistent with the "planet transits star A" scenario in \cite{Southworth2020}. The planetary radius uncertainty is confirmed as well with the TESS independent space observations and the orbital parameters are verified too. Overall, TESS confirmed an inflated giant with many kms of atmospheric scale height, suitable for transmission spectroscopy investigations. 

\begin{table*}
\begin{center}
\caption{Physical properties of \mbox{WASP-20 b}, as derived in this work with TESS data, in comparison to the previously published work of \cite{Anderson2015}. The dataset is in rather complete agreement with the work of \cite{Southworth2020} (see in text). In brackets, we present the derived values considering the third light contamination of the companion star.} 
\label{wasp20b_table}
\begin{tabular}{cccc}
\hline
\noalign{\smallskip}
&&This work - [L3 correction] &\cite{Anderson2015} \\
\noalign{\smallskip}
& Parameters[units] &Values $\pm 1\sigma$ &Values $\pm 1\sigma$ \\
\hline
\noalign{\smallskip}
&$\mathrm{T_c ~[BJD_{TDB}]}$ &$2455715.66591\pm 0.00034$ & $2455715.6564 \pm 0.0003$ \text{*}  \\ 
&$i$ [$^{\circ}$] &$85.82 \pm 0.16$ &$ 85.56\pm 0.22$ \\ 
&$\mathrm{a/R_s}$ &$9.50 \pm 0.17$& $9.29 \pm 0.23$ \\ 
&$\mathrm{R_p / R_s}$  &$0.09963 \pm 0.00047$ - [$0.1153 \pm0.0005$] &$ 0.10775\pm0.00102$\text{**} \\ 
& b &$0.691 \pm 0.028 $ &$ 0.718 \pm 0.018 $ \\ 
& $\mathrm{R_p ~ [R_J]}$&$1.38 \pm 0.04$ - [$1.43 \pm 0.05$] &$  1.458 \pm0.057$\\ 
&$\mathrm{ \rho_p ~[\rho_J]}$&$0.118\pm 0.012$ - [$0.108\pm0.013$] & $ 0.1006 \substack{+0.0131\\ -0.0099}$ \\ 
& $\mathrm{\log{g_p} ~[cgs]}$ &$2.585 \pm 0.027$ - [$2.46 \pm 0.03$] & $2.527 \pm0.036$\\ 
& $\mathrm{T_{eq} ~[K]}$ &$1362 \pm 26 $ &$ 1379 \pm 32$ \\ 
& $H$ ~[km] &$1340 \pm 90 $ - [$1790 \pm 123$] &-\\
\hline 
\end{tabular}
\end{center}
\begin{threeparttable}
\begin{tablenotes}
      
      \item * Converted from $2455715.65562\pm0.00028\, \mathrm{BJD_{UTC}}$
           \item ** Computed from the given $\mathrm{(R_p/R_s)}^2 = 0.01161\pm0.00022$ in \cite{Anderson2015}.
    \end{tablenotes}
 \end{threeparttable}
\end{table*} 

\subsection{WASP-108 b}
\mbox{WASP-108 b} was discovered and characterized in the work of \cite{Anderson2014}. The authors presented a bloated hot Jupiter of an inflated radius of $1.284\pm 0.047\,\mathrm{ R_J}$ which orbits a relatively bright ($V_{mag}=11.2$) F9 star on a $2.68$-days period. Recently, a companion star to \mbox{WASP-108} was reported in the work of \cite{Bohn2020}. 

Even though, this previous work of \cite{Anderson2014} appears only in the archive, eventually this planet is confirmed with independent space observations with TESS. Observing sectors 11, 37 and 38, provided a total of 25 transit light curves to analyze. Interestingly, our MCMC process, based on the emcee implementation, yielded a difference in $\mathrm{R_p/R_s}$ by more than $3\sigma$. However, the crowding metric indicated that there was a correction of the received light curves for a contamination of approximately $13\%$ by nearby sources, and this can be the main reason for the observed difference in the transit depth between our analysis and the previous work by \cite{Anderson2014}. We report also a later transit mid-time by $5.6$ minutes, while the orbital parameters are all in agreement with this previous investigation, within $1\sigma$. However, the derived planetary radius is consequently larger, since the third light contamination is taken into account, and supports now a less dense exoplanet scenario for \mbox{WASP-108 b}. The modeled TESS light curve is shown on the upper panel of Fig. \ref{wasp108b_corner}, while the derived parameters for this exoplanet are shown in the corner plot on the lower part of the figure. A direct comparison with the previously derived values of \cite{Anderson2014} is shown in Table \ref{wasp108b_table}{\ns}. 

\begin{table*}
\begin{center}
\caption{Physical properties of \mbox{WASP-108 b}, as derived in this work with TESS data, in comparison to the work of \cite{Anderson2014}.}
\label{wasp108b_table}
\begin{tabular}{cccc}
\hline
\noalign{\smallskip}
&&This work &\cite{Anderson2014} \\
\noalign{\smallskip}
& Parameters[units] & Values $\pm 1\sigma$ &Values $\pm 1\sigma$ \\
\hline
\noalign{\smallskip}
&$\mathrm{T_c ~[BJD_{TDB}]}$ &$ 2456413.79490\pm0.00018$ & $  2456413.79098 \pm 0.00015$  \text{*}\\ 
&$i$ [$^{\circ}$] &$89.2 \pm 0.7 $ &$  88.49 \pm 0.84$ \\ 
&$\mathrm{a/R_s}$ &$7.03 \pm 0.08$& $ 7.05 \pm 0.13$ \\ 
&$\mathrm{R_p / R_s}$   &$0.11165 \pm 0.00029$ &$0.10867 \pm 0.00069$ \text{**} \\ 
& b &$0.10 \pm 0.09$ &$  0.19 \pm 0.10 $ \\ 
& $\mathrm{R_p ~ [R_J]}$&$1.35 \pm 0.04 $ &$  1.284 \pm 0.047$\\ 
&$\mathrm{ \rho_p ~[\rho_J]}$&$0.36\pm 0.04$& $ 0.422 \pm 0.033$ \\ 
& $\mathrm{\log{g_p} ~[cgs]}$&$3.040 \pm 0.017$ & $  3.093 \pm 0.023$\\ 
& $\mathrm{T_{eq} ~[K]}$&$1589 \pm 33$ &$1590 \pm 36$ \\ 
&$H$ ~[km]&$547 \pm 26$ &- \\
\hline 
\end{tabular}
\end{center}
\begin{threeparttable}
\begin{tablenotes}
      
      \item * Converted from $2456413.79019\pm0.00015$ HJD
      \item ** Computed from the given $\mathrm{(R_p/R_s)}^2 = 0.01181\pm0.00015$ in \cite{Anderson2014}.
    \end{tablenotes}
 \end{threeparttable}
\end{table*}

\section{Discussion}
In the following paragraphs, we discuss the highlights of our analysis. More specifically, how the $\mathrm{R_p}$ (and the uncertainties on their measurements) derived from previous investigations and the measurements of $\mathrm{R_p}$ from our TESS analysis can be compared. Furthermore, we compare our newly derived b for these systems to the previously reported b values, and eventually we assess how these two quantities ($\mathrm{R_p}$ and b) are related. Better constrained b values, derived from a well acquired information on ingress and egress, place the planet to a more precise orbital configuration around its host. This consequently can yield a more robust $\mathrm{R_p}$ value, since the planetary radii are usually vulnerable towards the stellar limb darkening modeling process \citep{Alexoudi2020}.    

\subsection{Planetary radius refinement}
\begin{figure}[ht]
\begin{center}
\includegraphics[height=6.5cm]{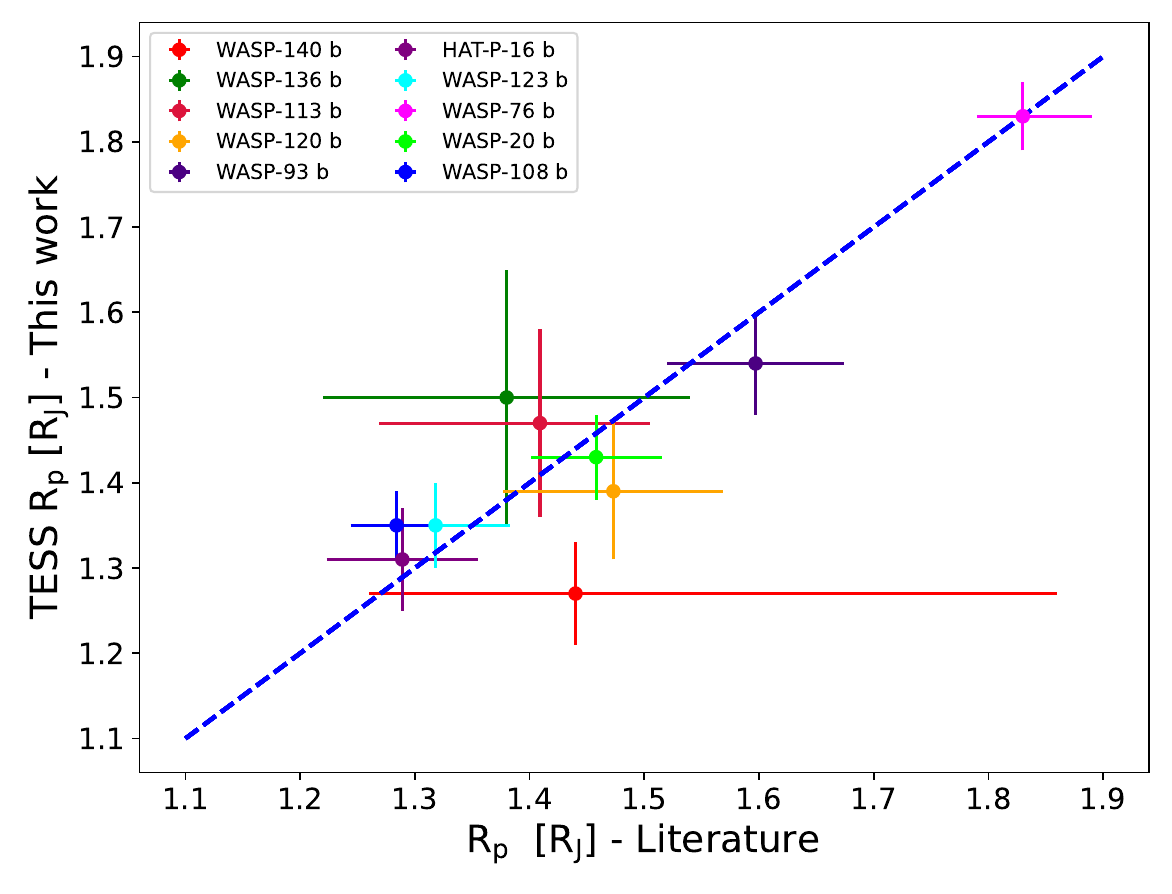}
\caption{The new planetary radii as derived from TESS observations for
each one of the exoplanets of our sample, in comparison to the previously published planetary radii values. The uncertainties on the radii are mostly the same within $1\sigma$. The exception is the uncertainty on the radius of \mbox{WASP-140 b},
which has been greatly improved by TESS.} 
\label{rp_rptess}
\end{center}
\end{figure}

For each one of the exoplanets of our sample, we derived their planetary radii through the analysis of TESS observations of single and multiple sectors. 
In Fig. \ref{rp_rptess}, we present the derived planetary radii with their uncertainties. The radii measurements are located approximately on the line of equality for $\mathrm{R_p}$ = $\mathrm{R_{p_{(TESS)}}}$. However, for \mbox{WASP-140 b} there is a significant decrease in the planetary radius by $12\%$ and its uncertainty is reduced by a factor of 3. This can be attributed to the fact that the transit analysis of the previously published work on this exoplanet concluded to a system with larger impact parameter value, hence a more grazing orbit. For the rest of exoplanets of the sample, the planetary radii agree within $ 1\sigma$ with the previous investigations. 
It is currently investigated by our group if TESS can improve to this extent the radii uncertainty of other exoplanets like \mbox{WASP-140 b}, regarding its grazing orbit around a host star of similar brightness. In our sample, \mbox{WASP-93} is similar to \mbox{WASP-140} in terms of a grazing orbit and similar host's apparent magnitude. However, for \mbox{WASP-93 b}, the radius uncertainty was confirmed by TESS, rather than improved. It is interesting to investigate if TESS can improve on the planetary radii and orbital parameters of similar systems to \mbox{WASP-140 b}, orbiting early K-type host stars. Interestingly, \mbox{WASP-177} \citep{Turner2019} is a K2 type host ($V_{mag}=12.3$) of a planet with a grazing orbit (b=0.98) and will be observed with TESS during sectors 42 and 55. \mbox{WASP-183 b} \citep{Turner2019}, which belongs also to a grazing system (b=0.92), orbits a slightly fainter star of G9/K0 type with $V_{mag}=12.76$ and will be observed by TESS during sectors 45 and 46. The photometric investigations of those two candidates might shed light whether there is a window in photometric observations where TESS in practice holds an advantage.  

\subsection{Impact Parameter refinement}
\begin{figure}[ht]
\begin{center}
\includegraphics[height=6.5cm]{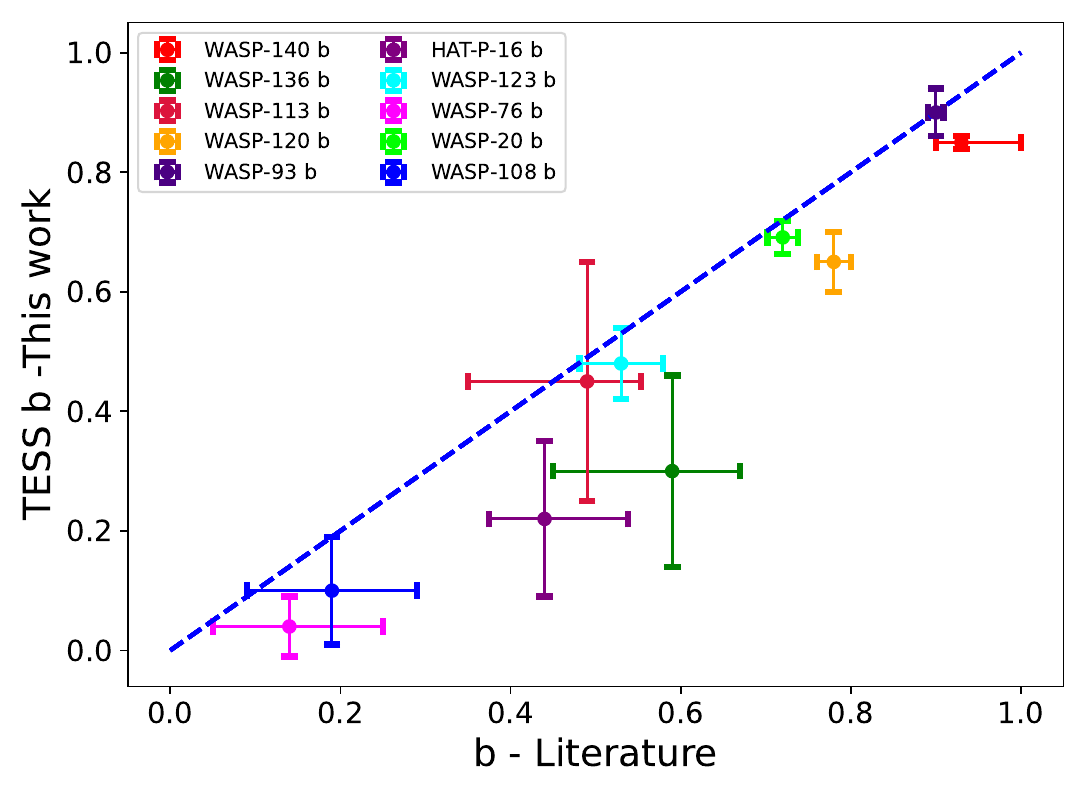}
\caption{Here, we depict a comparison between the b values from the literature, and the b measurements as derived from the analysis of high-cadence TESS data. TESS confirmed b within $2\sigma$ significance for all the exoplanets of our sample, except for \mbox{WASP-120 b}. The new b measurement for this exoplanet varies by more than $2\sigma$ from the previous estimation, since the TESS analysis refined significantly ($>2\sigma$) its orbital parameters of $i$ and $a/R_s$.} 
\label{impact_params}
\end{center}
\end{figure}

The analysis of the TESS datasets has shown that for some cases the orbital parameter refinement also resulted in a different impact parameter estimation. In Fig. \ref{impact_params}, we present the expected literature value for the impact parameter of each system and the derived TESS value. All the impact parameters derived with TESS are smaller (or equivalent) to the expected ones, likely due to the better continuous photometry provided by TESS. The ingress and egress of each system is monitored in high-quality data, and the orbital parameters $i$ and $a/R_s$ are better constrained. Overall, the TESS b values are in agreement with the previous investigations within $1\sigma$. Except the cases of \mbox{WASP-140 b}, \mbox{WASP-136 b}, \mbox{WASP-120 b} and \mbox{HAT-P-16 b}. For these, $i$ and $a/R_s$ deviate between $1-2.5\sigma$ from the previous researches, and this yields a b for these systems that deviates also accordingly.
For some systems of our sample, the eccentricity is not negligible. However, the non-zero eccentricity of \mbox{WASP-140 b} ($e=0.047 \pm0.004$), \mbox{WASP-120 b} ($e=0.059 \pm0.02$) and \mbox{HAT-P-16 b} ($e=0.036 \pm0.004)$, is expected to play a minor role in the derivation of the impact parameters, based on Eq.7 from \cite{Winn2010}.

\subsection{The relations between the planetary radii and the impact parameters}
The derived value of the radius of an exoplanet is strongly affected by the orbital set-up of the system. If the planet is transiting its host star centrally (~b=0) or over the limbs (~b=1), this plays a major role in the correct estimation of the transit depth \citep{Alexoudi2020}. As we saw previously, b depends on $i$ and $a/R_s$. This information on the orbital parameters is usually acquired from a well-defined ingress and egress in the received light curve. For specific ground-based observations that the analysis is solely based on  light curves from partial transits or really noisy datasets, this might yield poorly constrained orbital parameters. However, TESS with a 27-days uninterrupted photometry, confirmed b for most of the planets in our sample, while it also provided a better constrained b for the ones that their orbital parameters were refined within $ 2\sigma$ from the previous investigations. 

\begin{figure}[ht]
\begin{center}
\includegraphics[height=6.5cm]{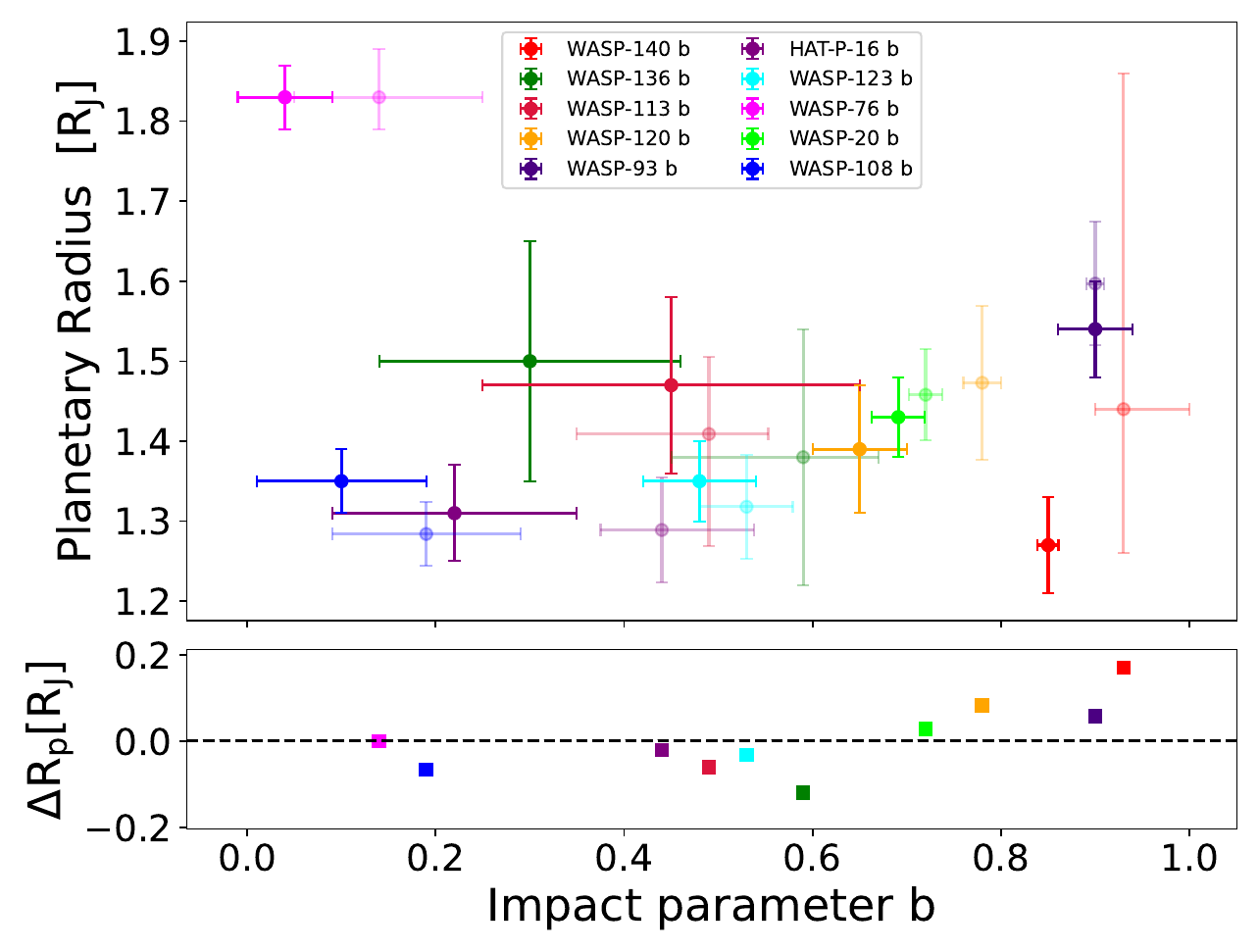}
\caption{In the upper panel, we show the planetary radii derived in this work, in comparison to the impact parameter of these systems. In fainter colors, we present the literature measurements, while in vivid colors we depict the measurements as derived from the analysis of high-cadence TESS data. \mbox{WASP-140 b},\mbox{WASP-136 b}, \mbox{WASP-120 b} and \mbox{HAT-P-16 b} are not in a complete agreement with the previous investigations. On the lower panel, we present the difference between the planetary radius reported in the literature and the planetary radius derived from TESS observations (colored squares). We plotted this difference versus the literature b value. For b>0.6, the $\mathrm{R_p}$ values appear slightly overestimated by the ground-based investigations.}
\label{rp_impact}
\end{center}
\end{figure}

In the upper panel, in Fig. \ref{rp_impact}, we present the previous investigations of the $\mathrm{R_p}$ and b measurements of these systems in fainter colors, while in vivid colors we depict the results as derived in this work using TESS. At the lower panel, it is noticeable that for systems of b>0.6 the planetary radii have been overestimated by the previous investigations, while for centrally transiting systems the published radii were slightly underestimated. This is an indication that poorly constrained grazing systems likely yield an overestimation of planetary radii. However, our sample is small to verify this. The importance of this finding is currently under investigation by our team, using a rich sample of grazing systems observed with TESS.

\section{Summary and Conclusions}

In this work, we employed a small sample of inflated hot giants and using high-quality uninterrupted datasets, provided by TESS, we aimed to improve on their planetary radius uncertainty. For this purpose, we chose exoplanets that have the largest reported uncertainty on their planetary radii, in the literature. The precise measurement of the planetary radius can help the evolutionary models to predict the presence or the absence of a planetary core and the interior energy of the planet \citep{Bodenheimer2003}. 

First, we obtained the publicly available transit light curves from each system and then we fit those light curves using a detrending and a transit model. Through a MCMC approach, using emcee, we provided best fit values for each model that was applied on the light curves. Then we used these best fit values to derive the new planetary parameters for each system.  

The results have shown that only \mbox{WASP-140 b} was benefited significantly from the TESS investigation with respect to its planetary radius uncertainty. Nevertheless, by using TESS's unprecedented precision, we were able to report the planetary radii refinement/confirmation of a sample of inflated hot giant exoplanets. The high precision of TESS not only yielded refined planetary parameters for \mbox{WASP-140 b}, it also improved the orbital parameters of \mbox{WASP-120 b}. For the latter, the large amount of continuous high-quality data, spread over four sectors of TESS observations, was the reason to derive better constrained orbital parameters for this system. Another outcome of our investigation is the clarification on the discrepancy regarding the planetary radius of \mbox{WASP-93 b}, between the independent investigations in  \cite{Hay2016} and \cite{Gajdos2019}. For all the other exoplanets of our sample, the derived parameters are in agreement with the previous investigations (within $1-2\sigma$). Last but not least, we report an indication that the ground-based investigations likely have overestimated the $\mathrm{R_p}$ value for systems with b>0.6. This needs to be confirmed with the analysis of high-quality TESS datasets of a rich sample of (near-)grazing systems. 

For most of the cases, TESS confirmed the previous investigations demonstrating that excellent photometric studies can be achieved with small telescopes of the class of $1$m  (e.g. 1.2m STELLA\footnote{\href{stella.aip.de}{stella.aip.de} } \citep{Strassmeier2004,Strassmeier2010} or the 1.23m CAHA\footnote{\href{https://www.caha.es/CAHA/Telescopes/1.2m.html}{www.caha.es/CAHA/Telescopes/1.2m.htm}}), that can reach photometric sensitivity of \mbox{1 mmag} (e.g. \cite{Ciceri2013, Mallonn2015}) or even lower in some cases \citep{Mallonn2021}. However, the correct parameterization from the ground is sometimes vulnerable, due to the large point-to-point scatter of the data at the transit light curve, or due to incomplete transit datasets that lack the information on ingress and egress. For these investigations, TESS, with uninterrupted (exception is the data downloading time) and high-quality light curves, can contribute as a complementary tool to the ground-based investigations, in order to verify, or to refine the parameters of exoplanetary systems with independent measurements from space. Moreover, TESS observations could be combined with ground-based observations delivering high-fidelity datasets. One option is to try to combine TESS's precision with high-quality photometry from ground-based instruments in order to construct a transmission spectrum for atmospheric characterization \citep{Yip2021}. Eventually, future instrumentation (e.g.\ JWST \citep{Gardner2006}, PLATO \citep{Rauer2014}) is expected to improve on the exoplanetary parameters tremendously. Till then, TESS will hold the fort by providing photometric data of the highest quality, available to the community for further researches and investigations on exoplanetary systems.

%===============================================================================
%    ACKNOWLEDGEMENTS
%===============================================================================

\section{Acknowledgments}
We would like to thank Klaus Strassmeier, Matthias Mallonn, Katja Poppenhaeger, Ekaterina Dineva and Engin Keles, for useful discussions, questions, comments and suggestions that helped to improve this work significantly. We sincerely appreciate the useful comments and suggestions from the anonymous reviewer, which enriched the quality of the manuscript. XA thanks Ioannis Mitsos for fruitful discussions regarding code structure and optimization. 
This paper includes data collected by the TESS mission, which are
publicly available from the Mikulski Archive for Space Telescopes (MAST). Funding for the TESS mission is provided by the NASA Explorer Program.
This research made use of Lightkurve, a Python package for Kepler and TESS data analysis (Lightkurve Collaboration, 2018).
XA is grateful for the financial support from the Potsdam Graduate School (PoGS) in the form of a doctoral scholarship.

%===============================================================================
%    BIBLIOGRAPHY
%===============================================================================

\bibliographystyle{Wiley-ASNA}
\bibliography{ an-jour, mybib}

%\end{document}
%===============================================================================
%    APPENDIX
%===============================================================================

\appendix
\section{Best fit transit models and corner plots for all the fit parameters.}

\begin{figure*}[ht]
\begin{center}
\includegraphics[height=6cm]{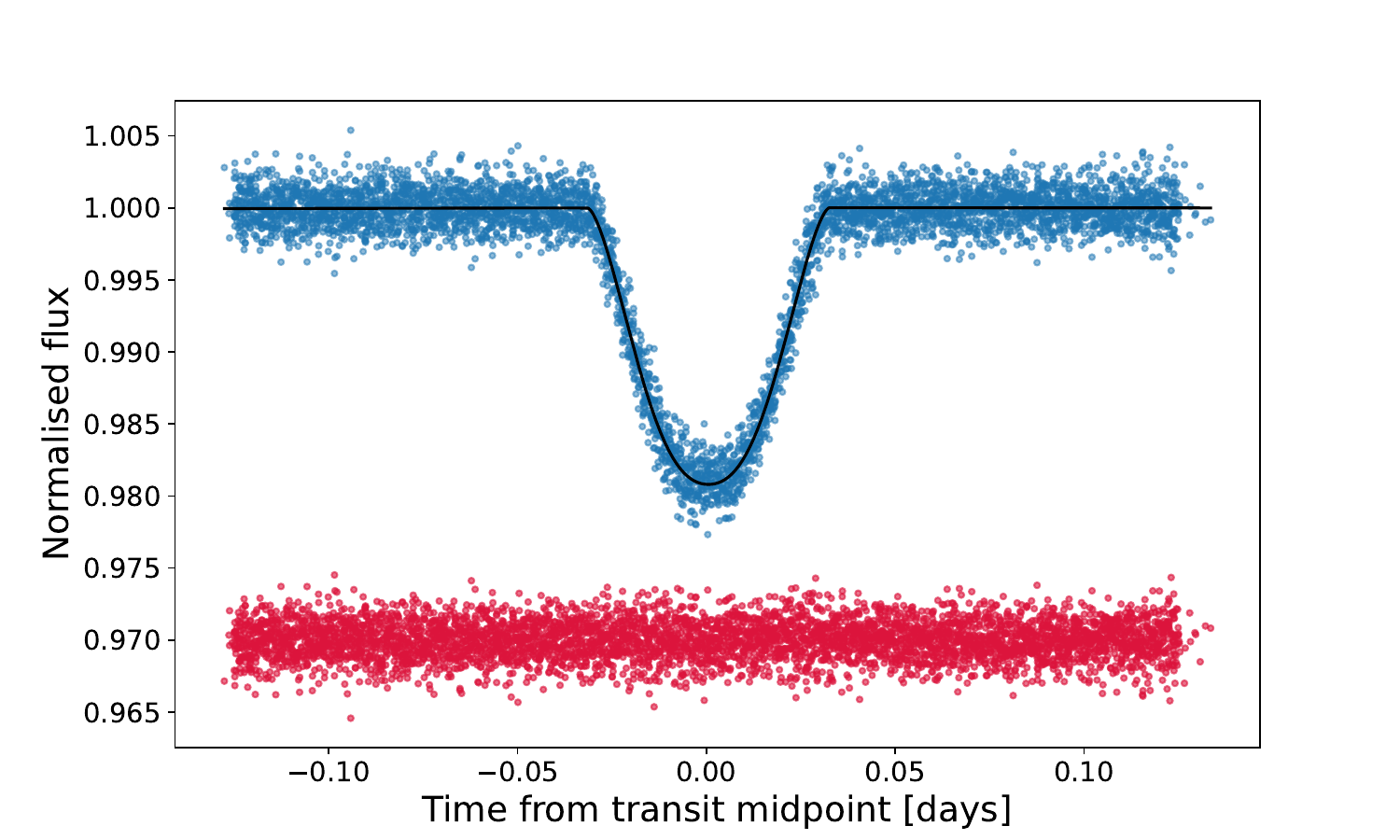}
\end{center}
\includegraphics[height=15cm]{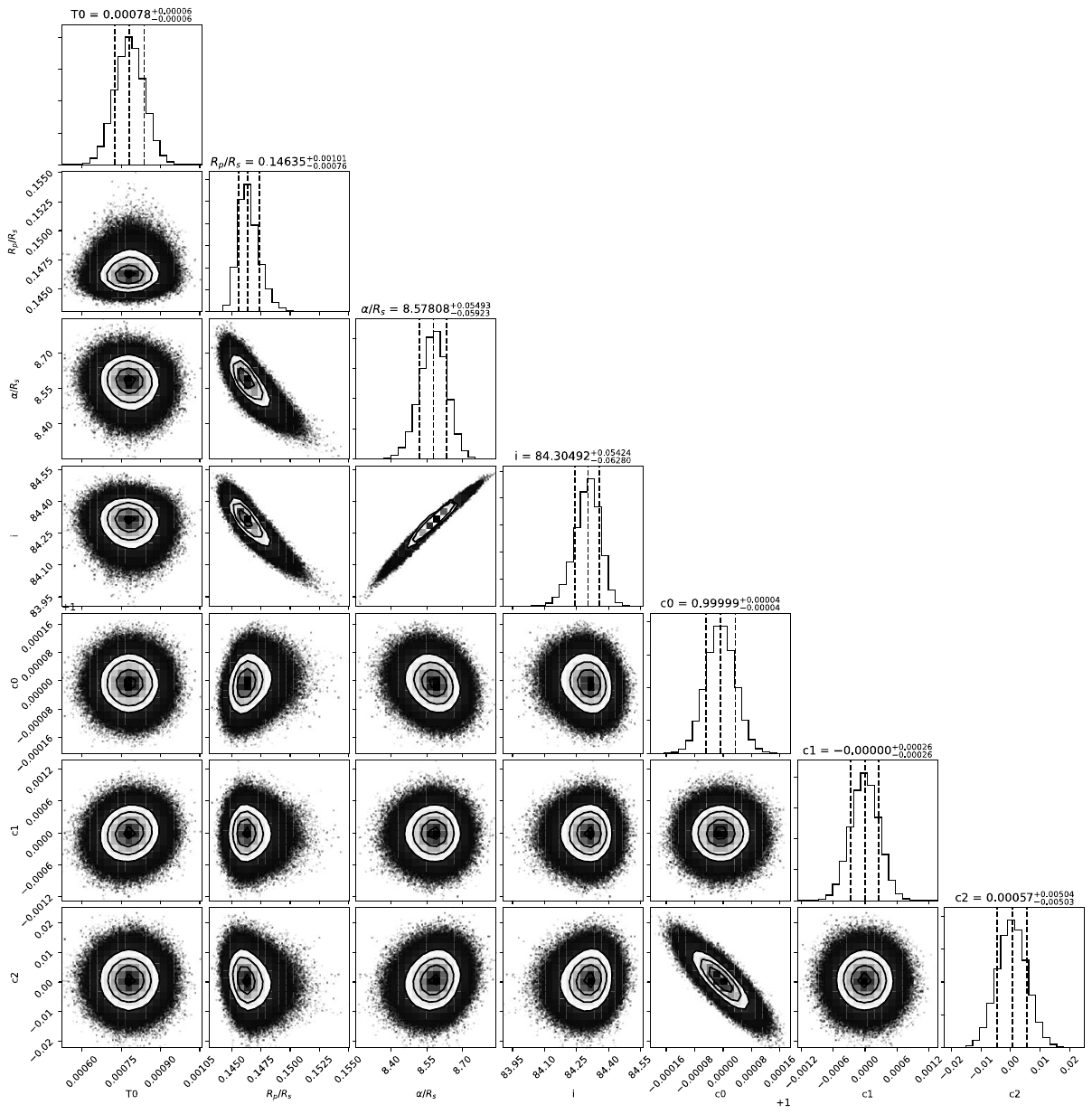}
\caption{In the upper panel, we show the folded TESS light curves of \mbox{WASP-140 b} (blue dots) along with the best fit transit model (black solid line). The residuals are presented with red dots and an offset for clarity. In the lower panel, we show the corner plot of the best fit parameters. It is the 2D projection of the sample plotted in a way to show covariance between the parameters.}
\label{wasp140b_corner}
\end{figure*}

\begin{figure*}[ht]
\begin{center}
\includegraphics[height=6cm]{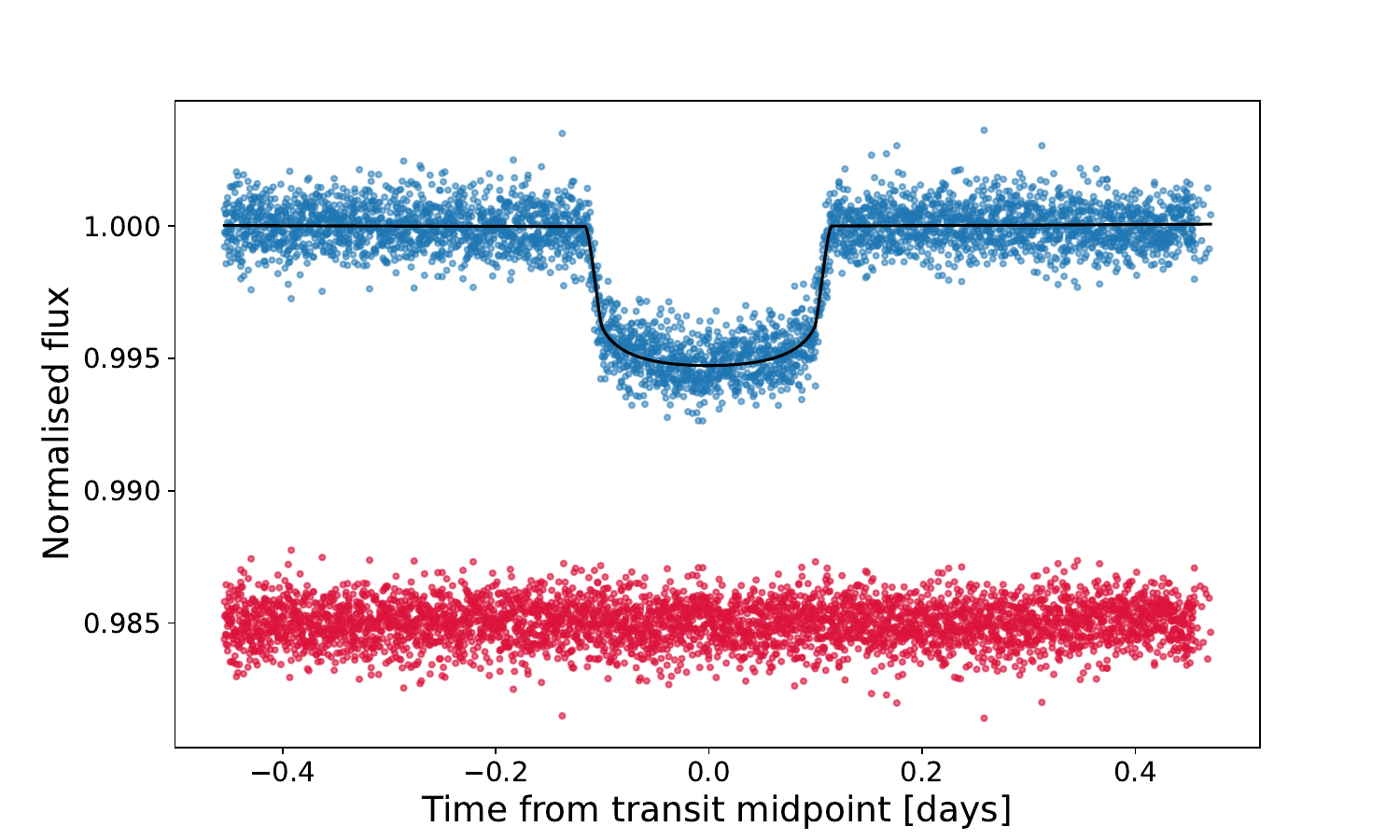}
\end{center}
\includegraphics[height=15cm]{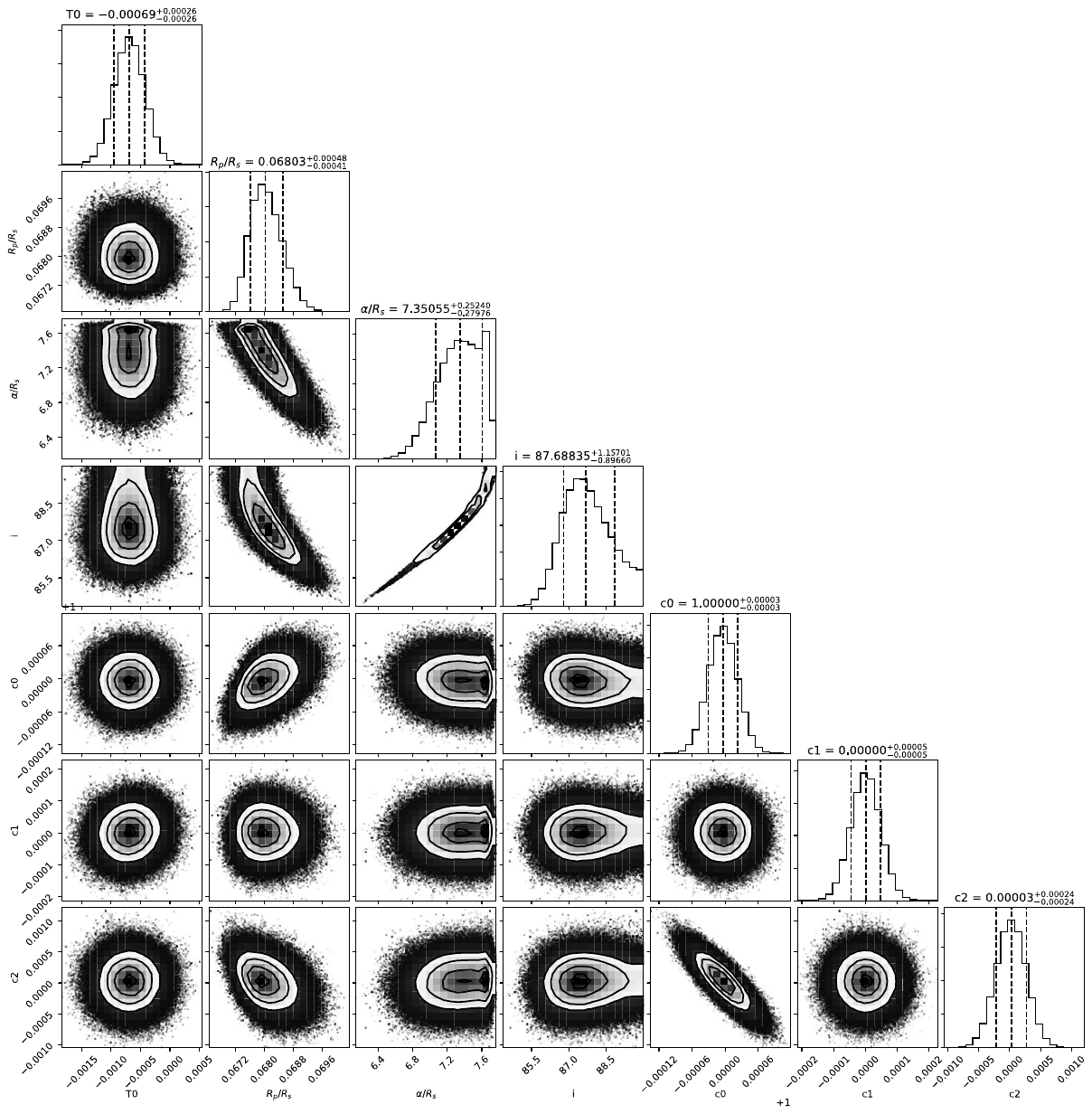}
\caption{The same as Fig. \ref{wasp140b_corner}, but for the exoplanet \mbox{WASP-136 b}.}
\label{wasp136b_corner}
\end{figure*}

\begin{figure*}[ht]
\begin{center}
\includegraphics[height=6cm]{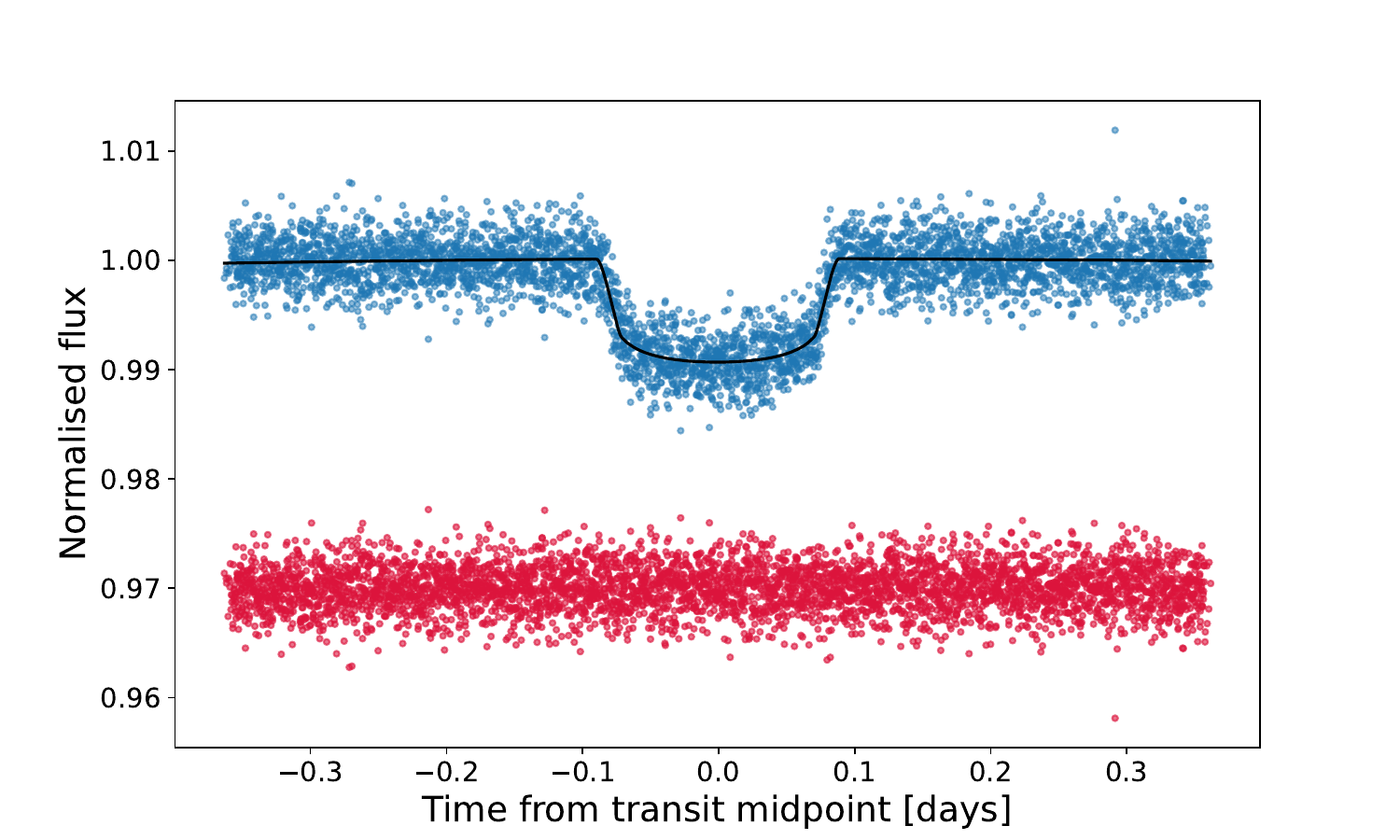}
\end{center}
\includegraphics[height=15cm]{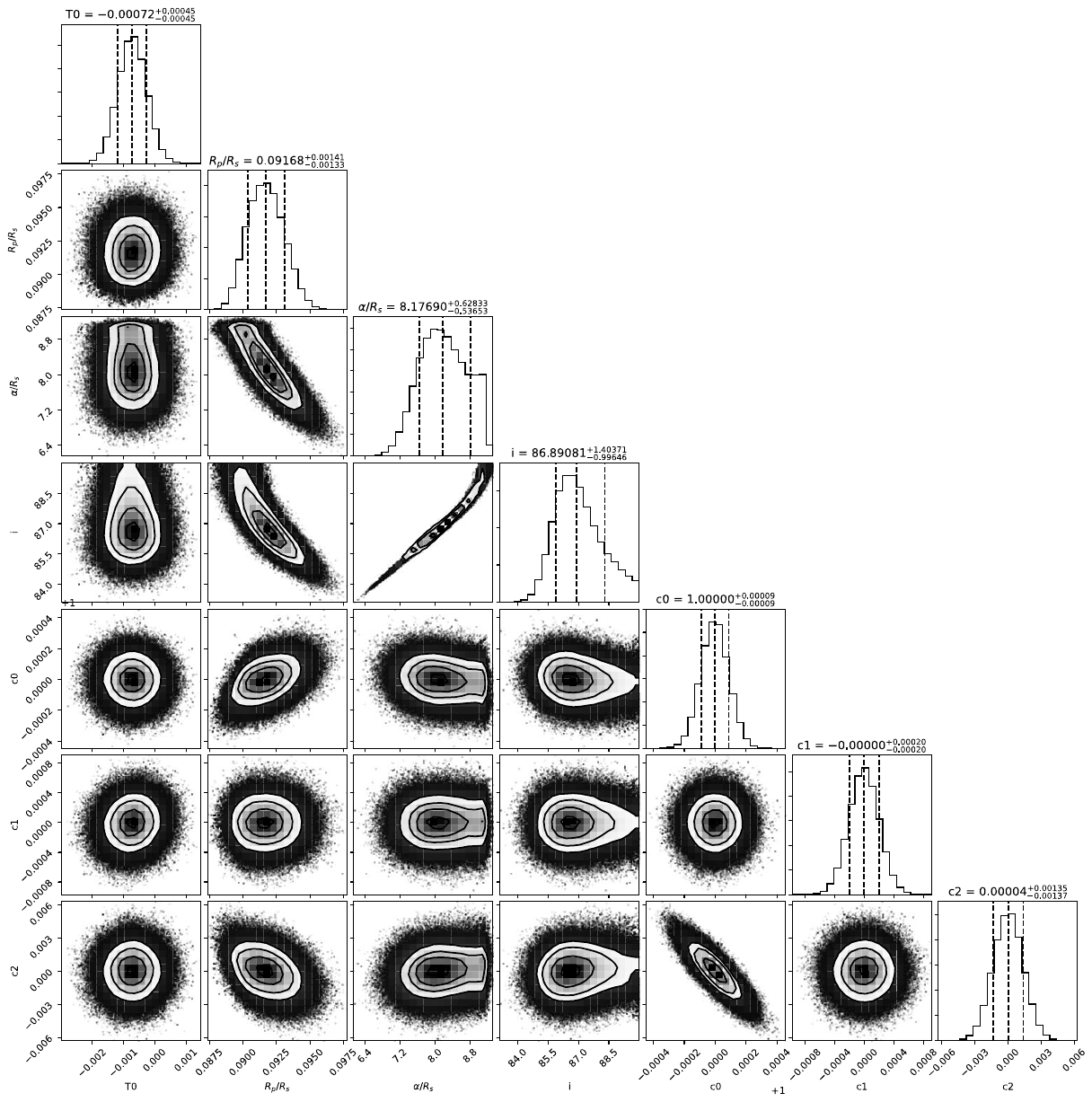}
\caption{The same as Fig. \ref{wasp140b_corner}, but for the exoplanet \mbox{WASP-113 b}.}
\label{wasp113b_corner}
\end{figure*}

\begin{figure*}[ht]
\begin{center}
\includegraphics[height=6cm]{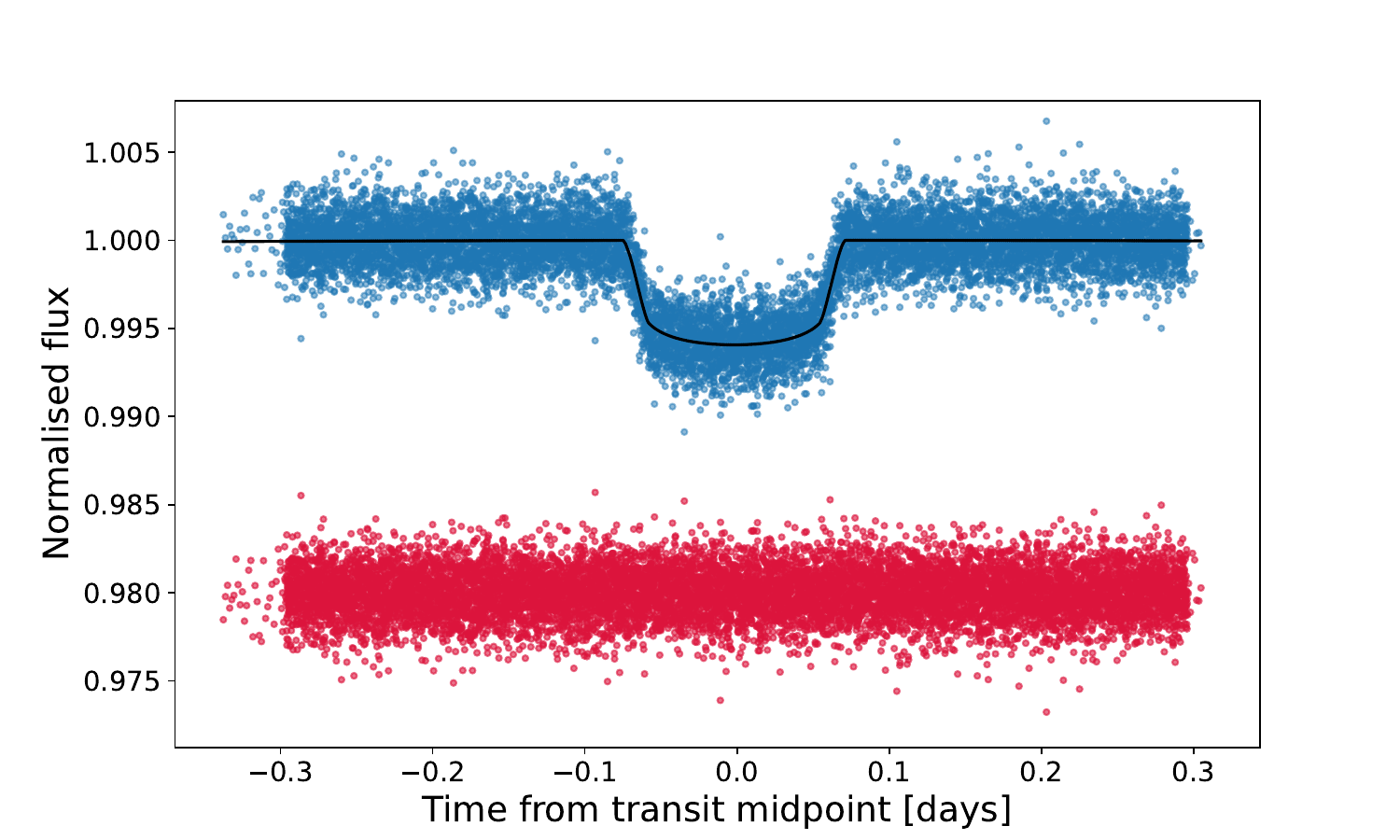}
\end{center}
\includegraphics[height=15cm]{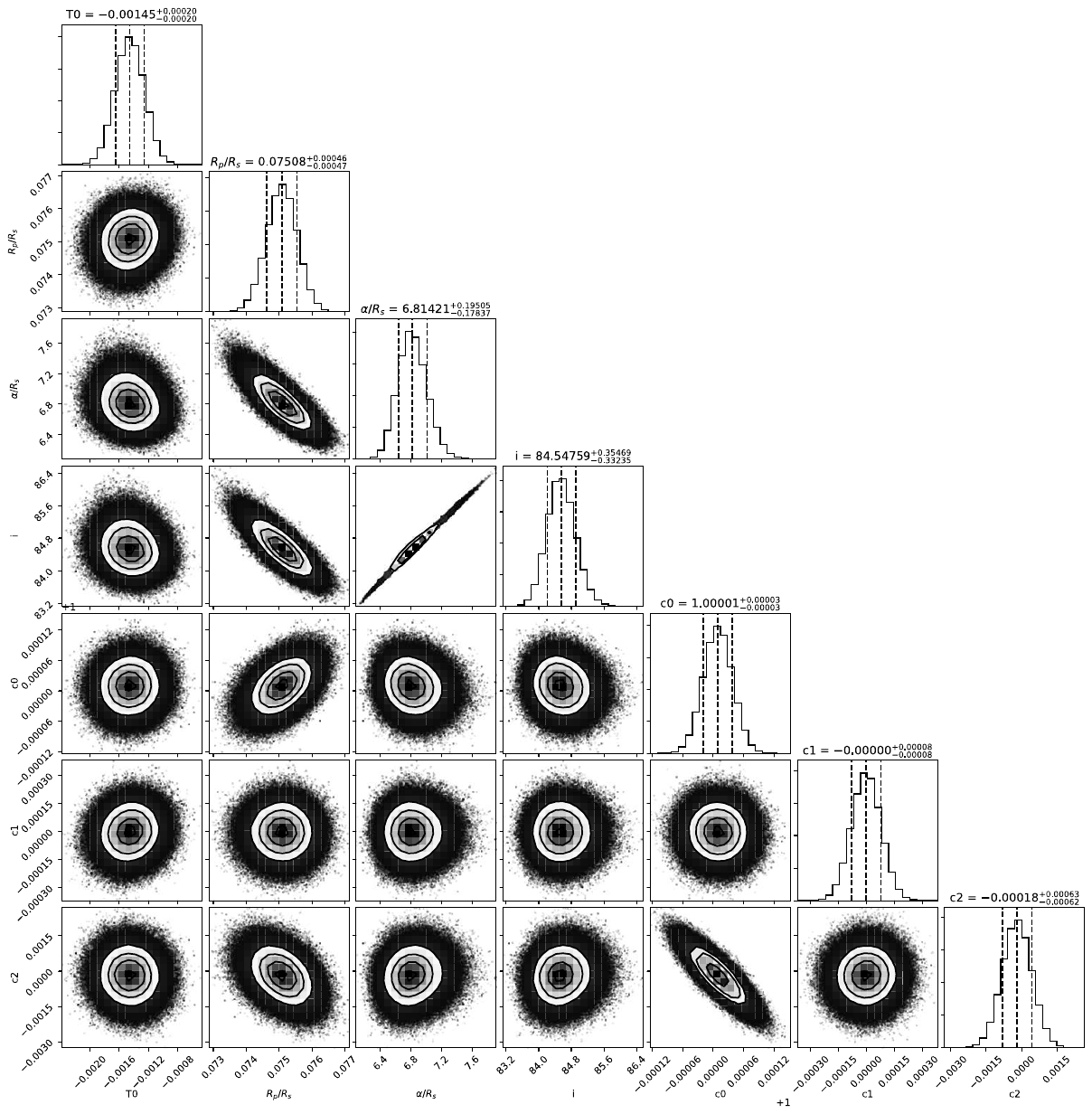}
\caption{The same as Fig. \ref{wasp140b_corner}, but for the exoplanet \mbox{WASP-120 b}.}
\label{wasp120b_corner}
\end{figure*}

\begin{figure*}[ht]
\begin{center}
\includegraphics[height=6cm]{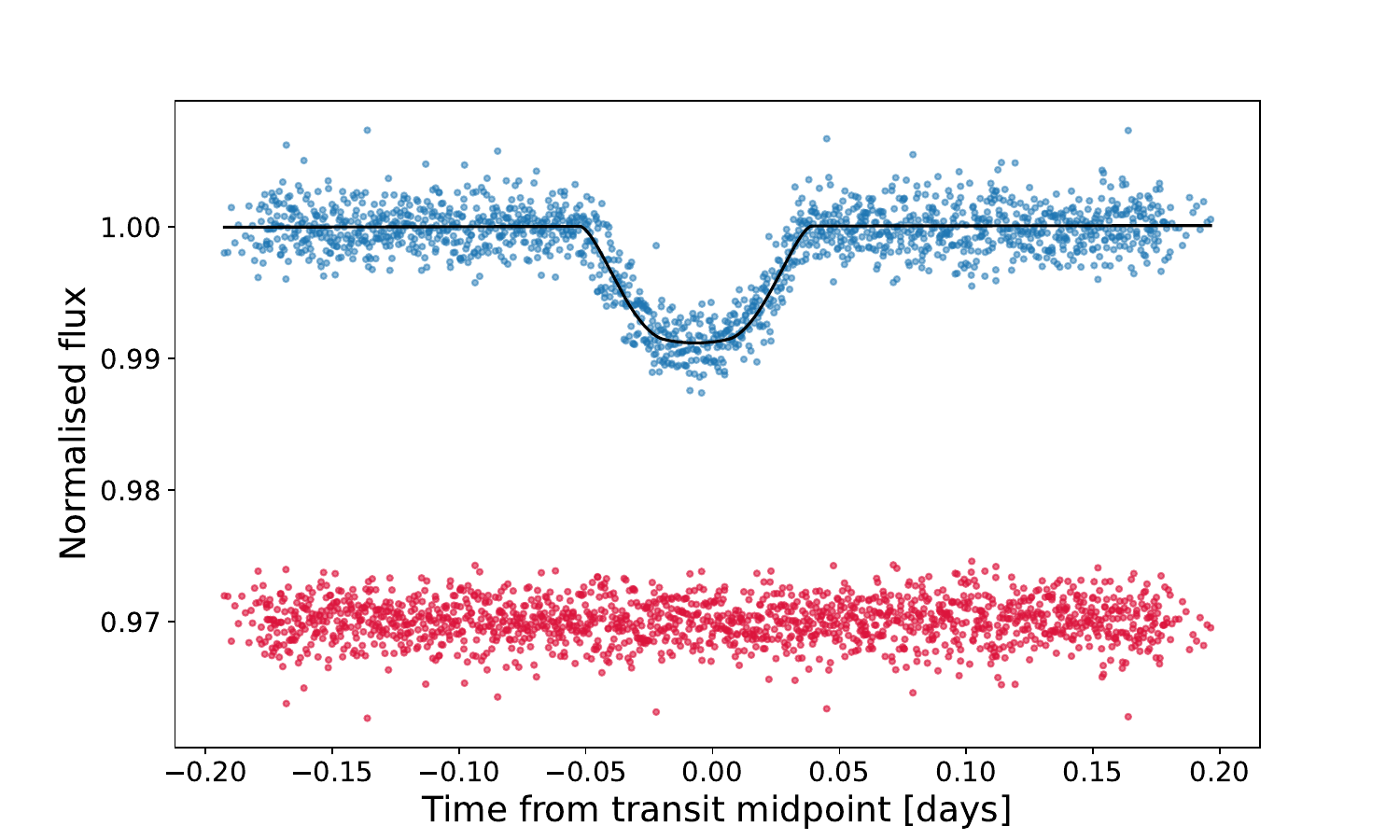}
\end{center}
\includegraphics[height=15cm]{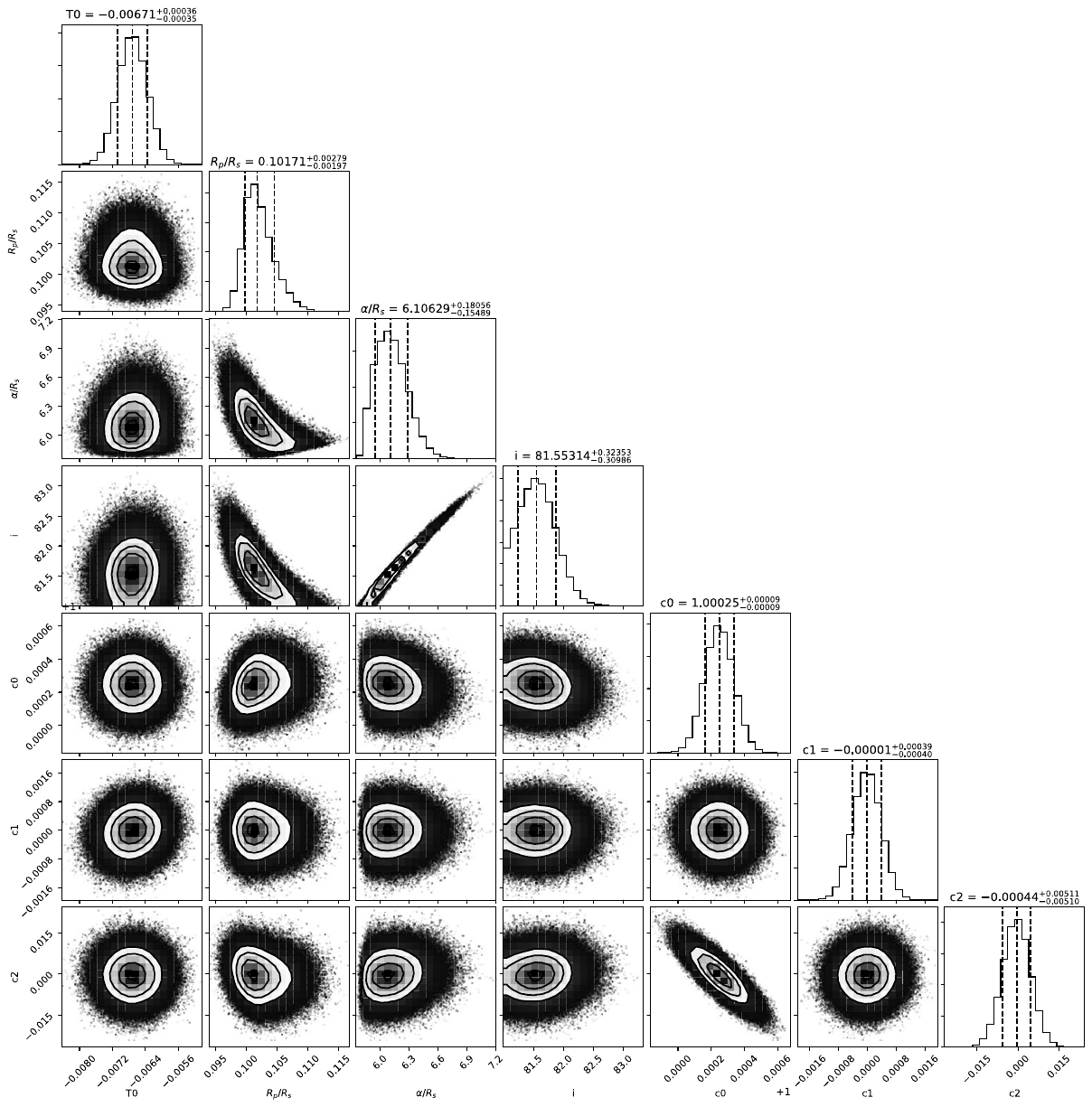}
\caption{The same as Fig. \ref{wasp140b_corner}, but for the exoplanet \mbox{WASP-93 b}.}
\label{wasp93b_corner}
\end{figure*}

\begin{figure*}[ht]
\begin{center}
\includegraphics[height=6cm]{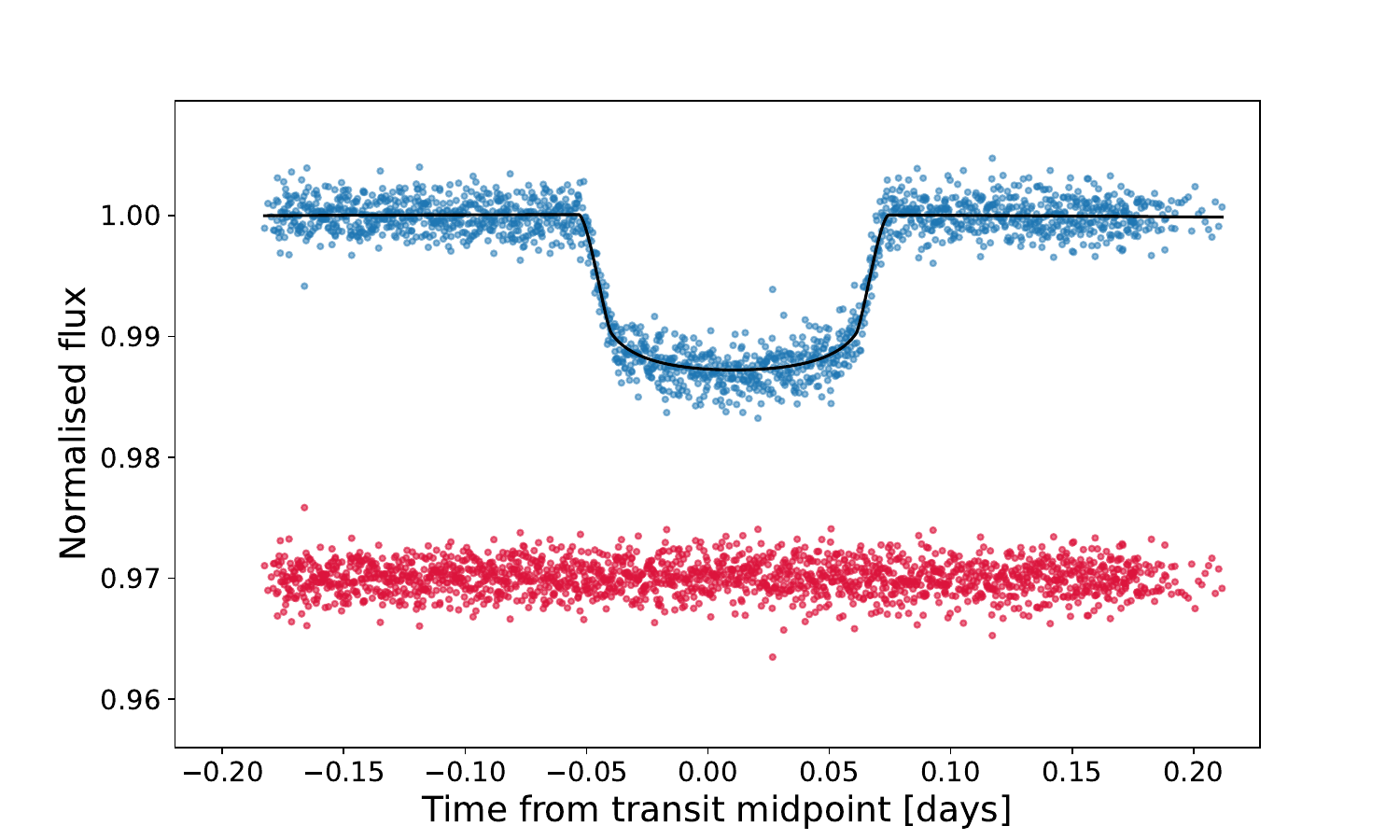}
\end{center}
\includegraphics[height=15cm]{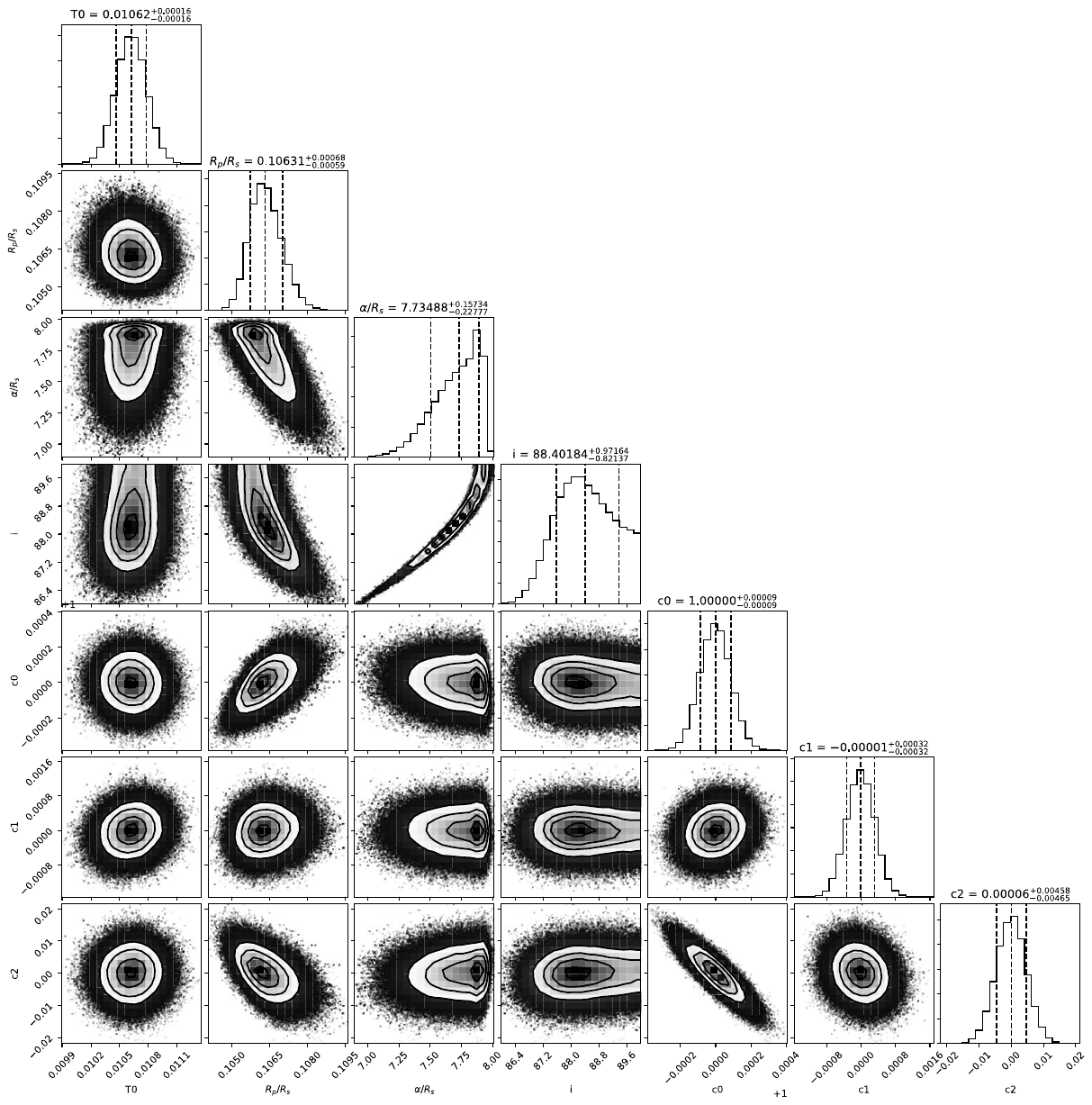}
\caption{The same as Fig. \ref{wasp140b_corner}, but for the exoplanet \mbox{HAT-P-16 b}.}
\label{hatp16b_corner}
\end{figure*}

\begin{figure*}[ht]
\begin{center}
\includegraphics[height=5cm]{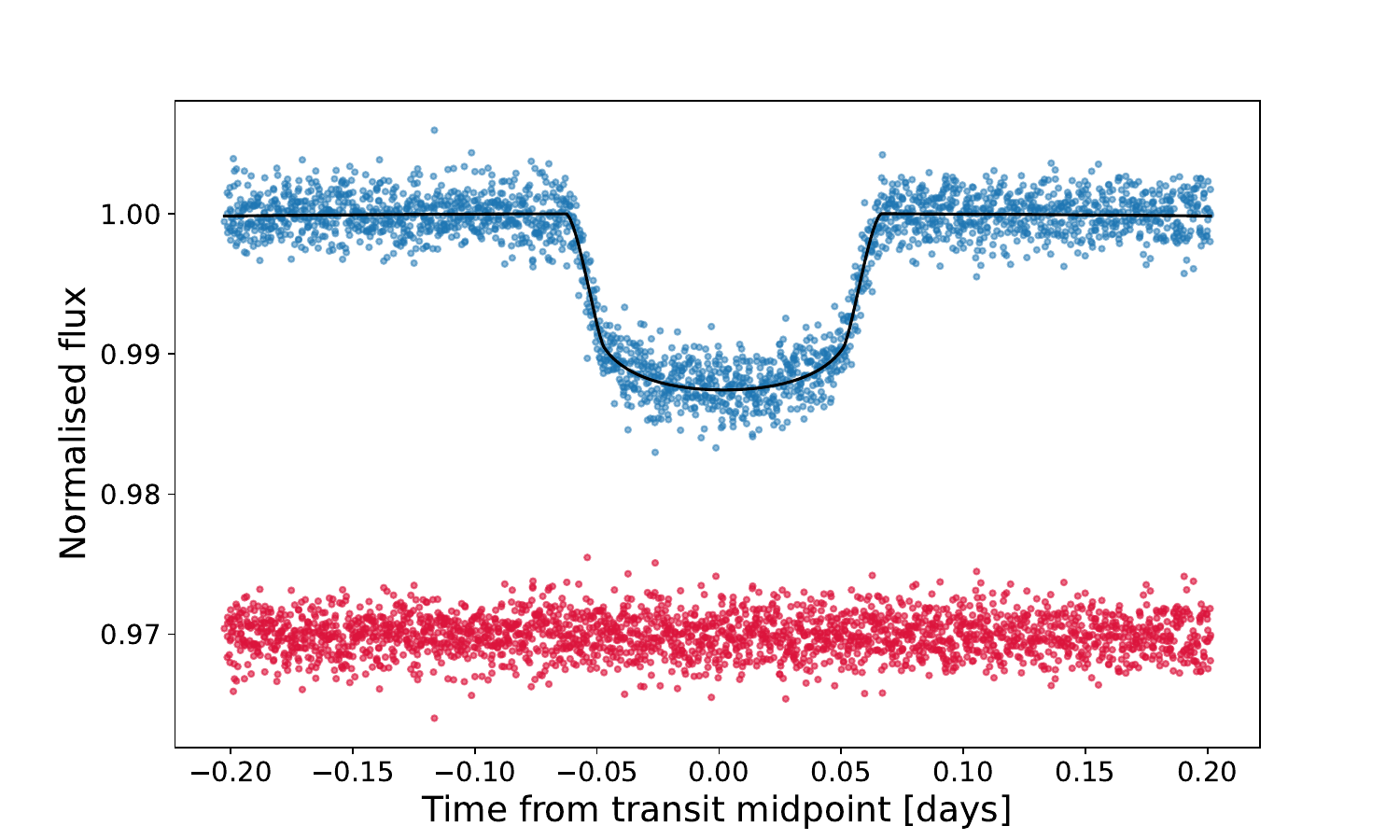}
\includegraphics[height=5cm]{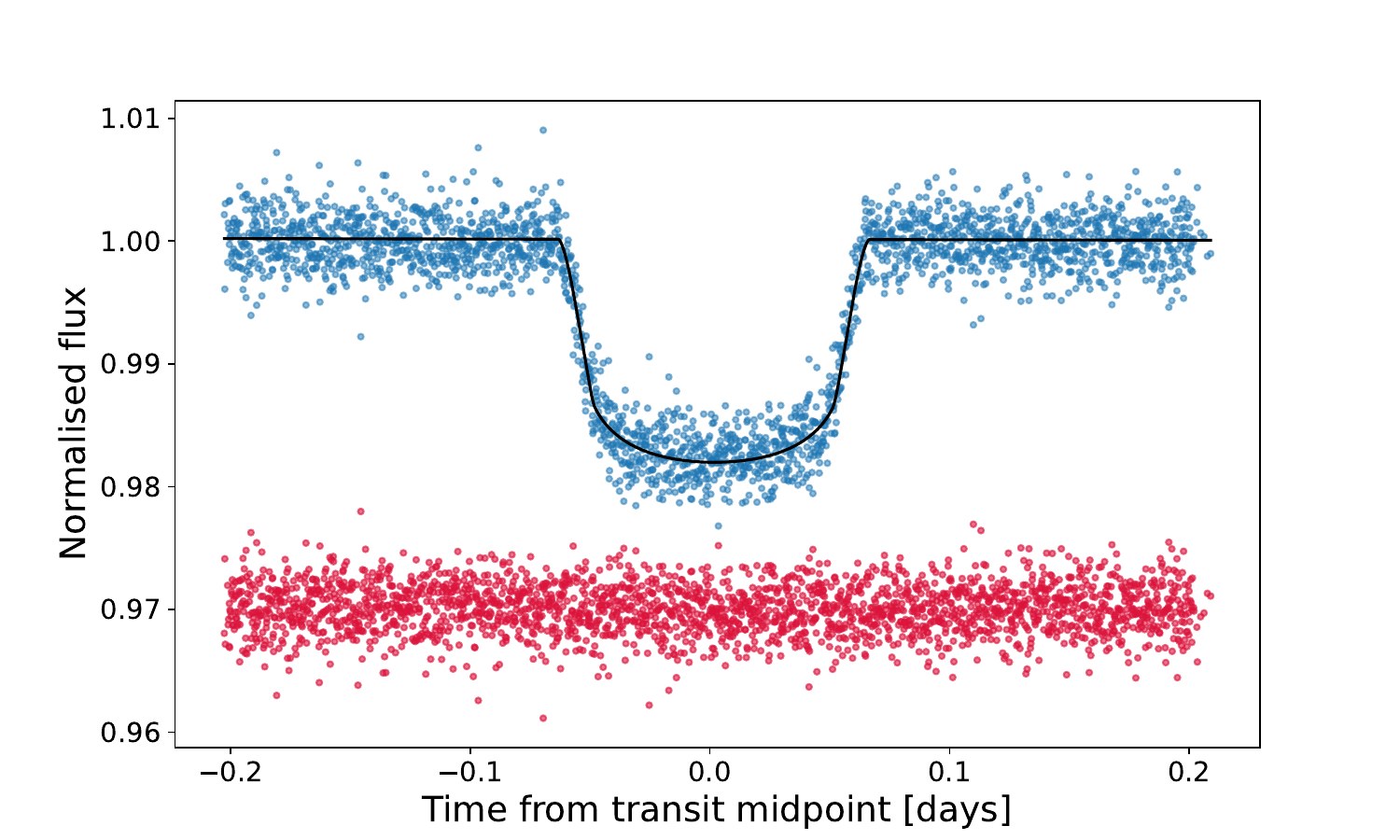}
\end{center}
\includegraphics[height=15cm]{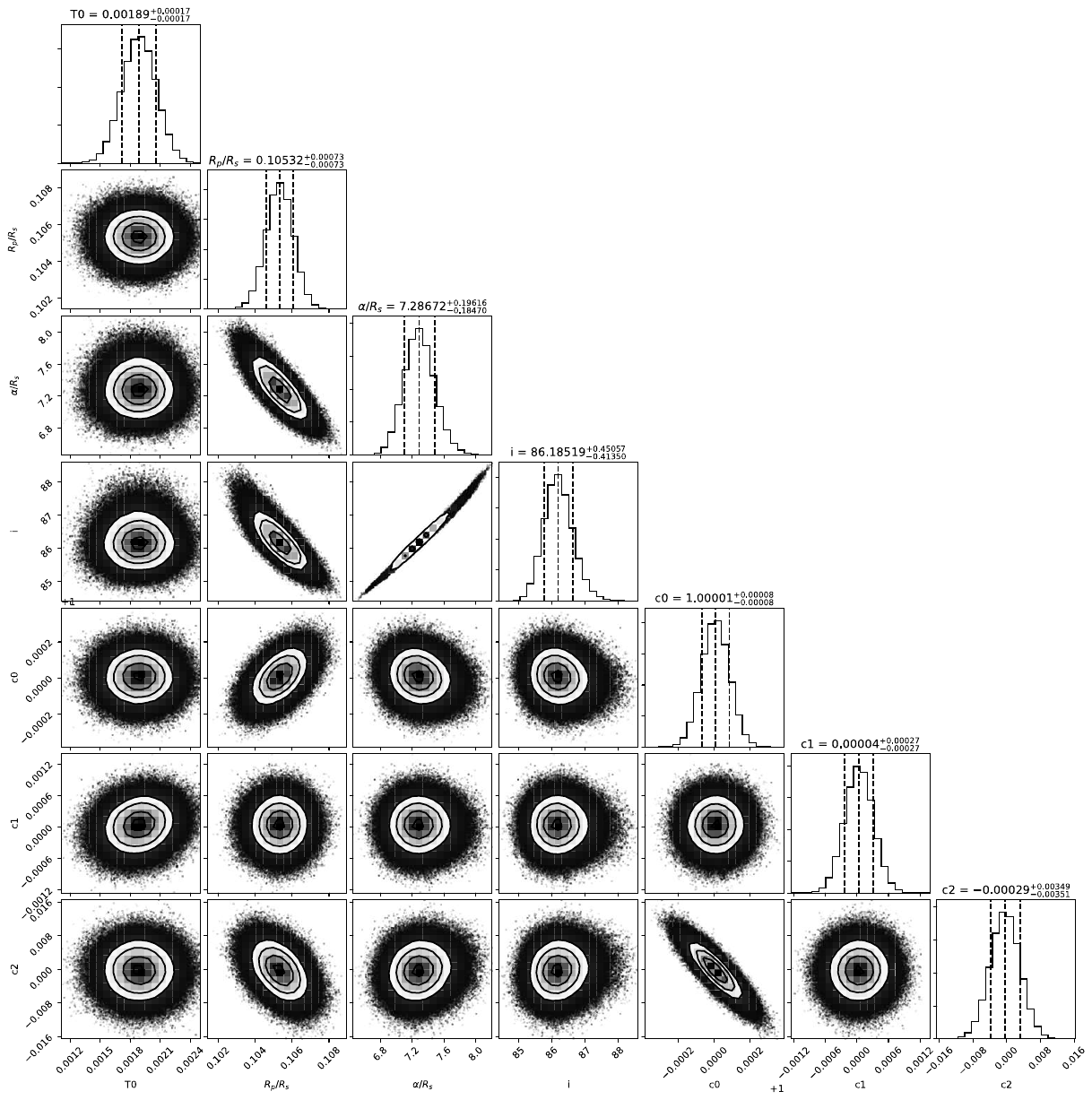}
\caption{Best fit transit model for the TESS light curves of \mbox{WASP-123 b}. On the upper left panel there are observations from sector 27 and on the upper right panel from sector 13. On the lower panel, the corner plot of the best fit parameters for \mbox{WASP-123 b}.}
\label{wasp123b_corner}
\end{figure*}

\begin{figure*}[ht]
\begin{center}
\includegraphics[height=6cm]{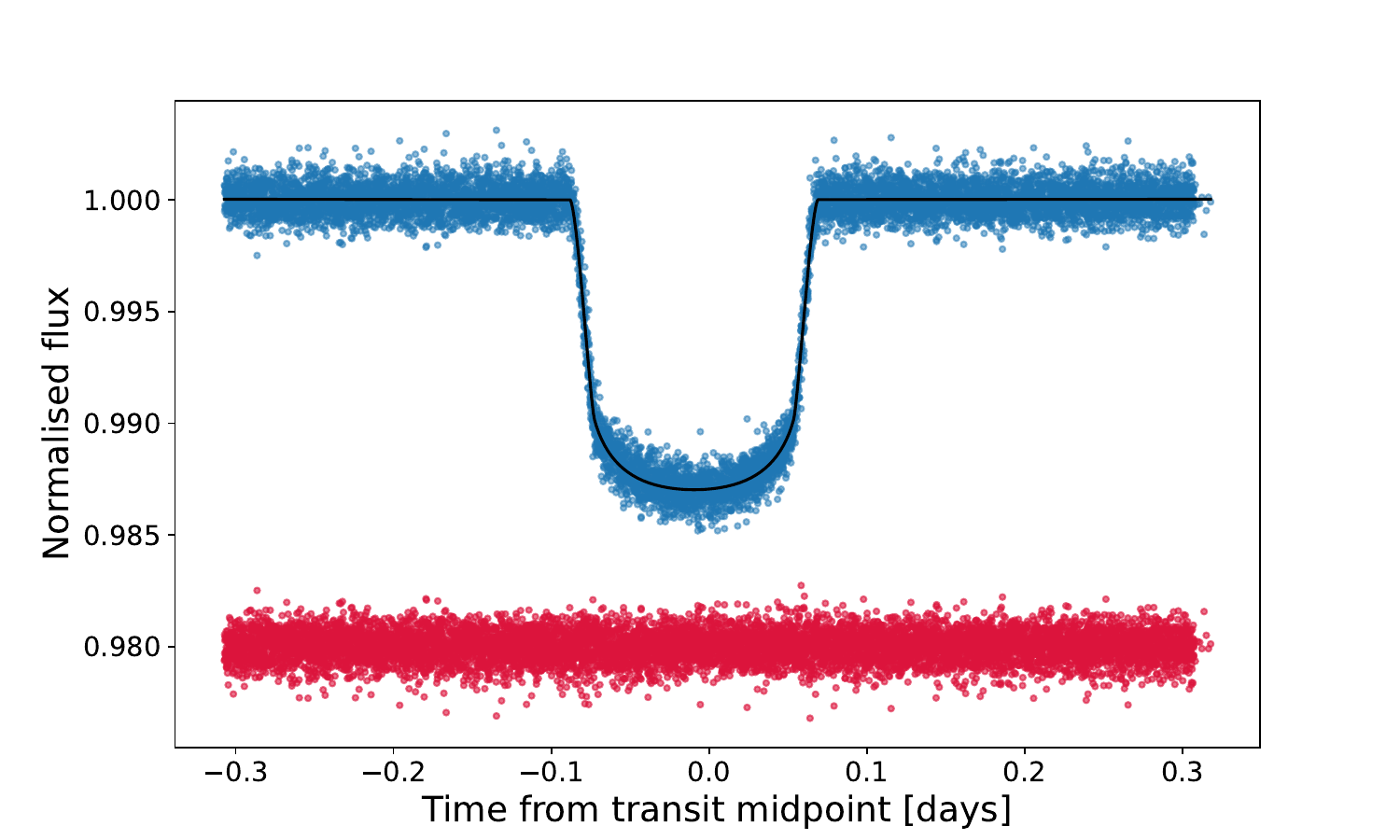}
\end{center}
\includegraphics[height=15cm]{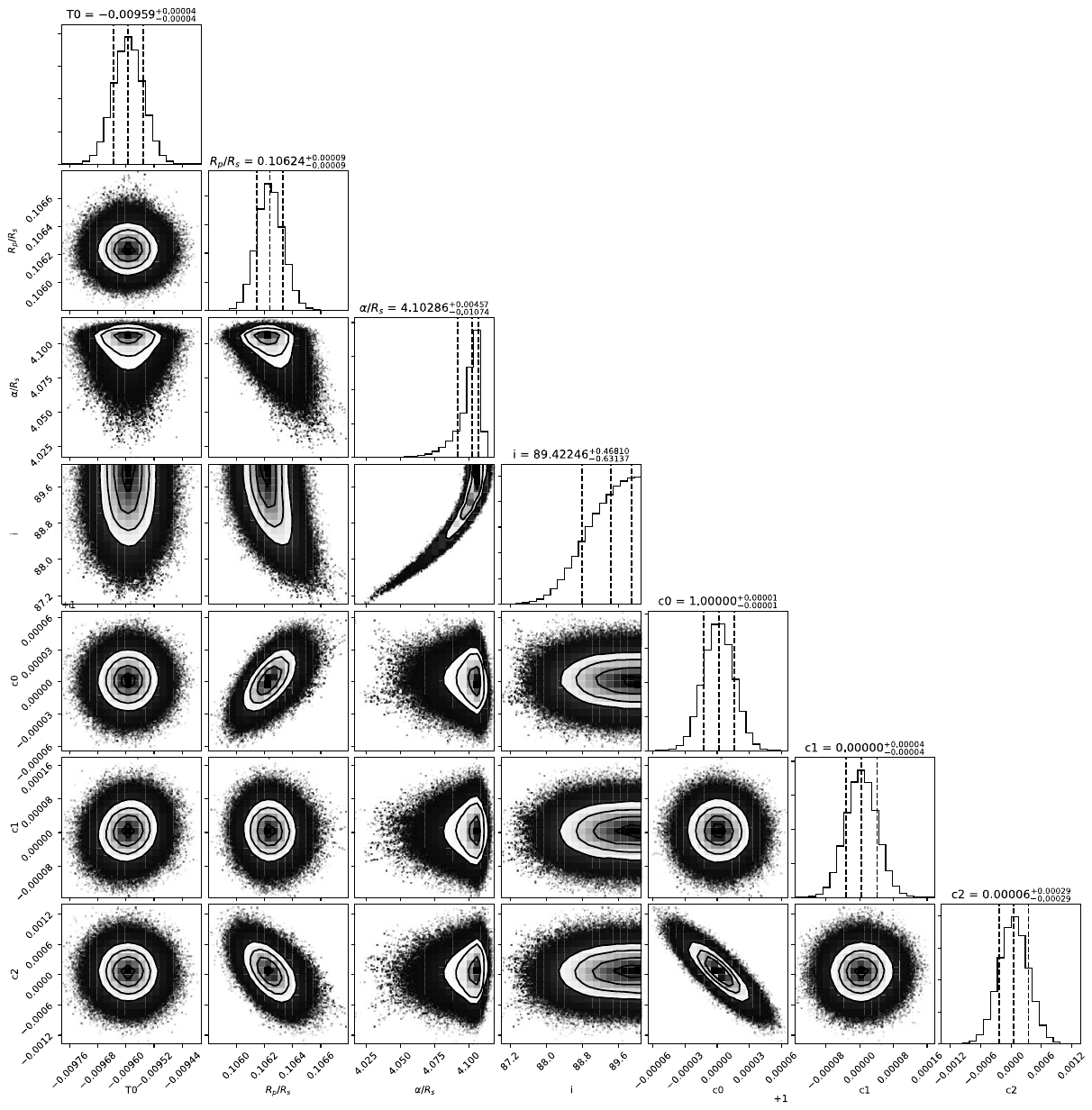}
\caption{The same as Fig. \ref{wasp140b_corner}, but for the exoplanet \mbox{WASP-76 b}.}
\label{wasp76b_corner}
\end{figure*}

\begin{figure*}[ht]
\begin{center}
\includegraphics[height=6cm]{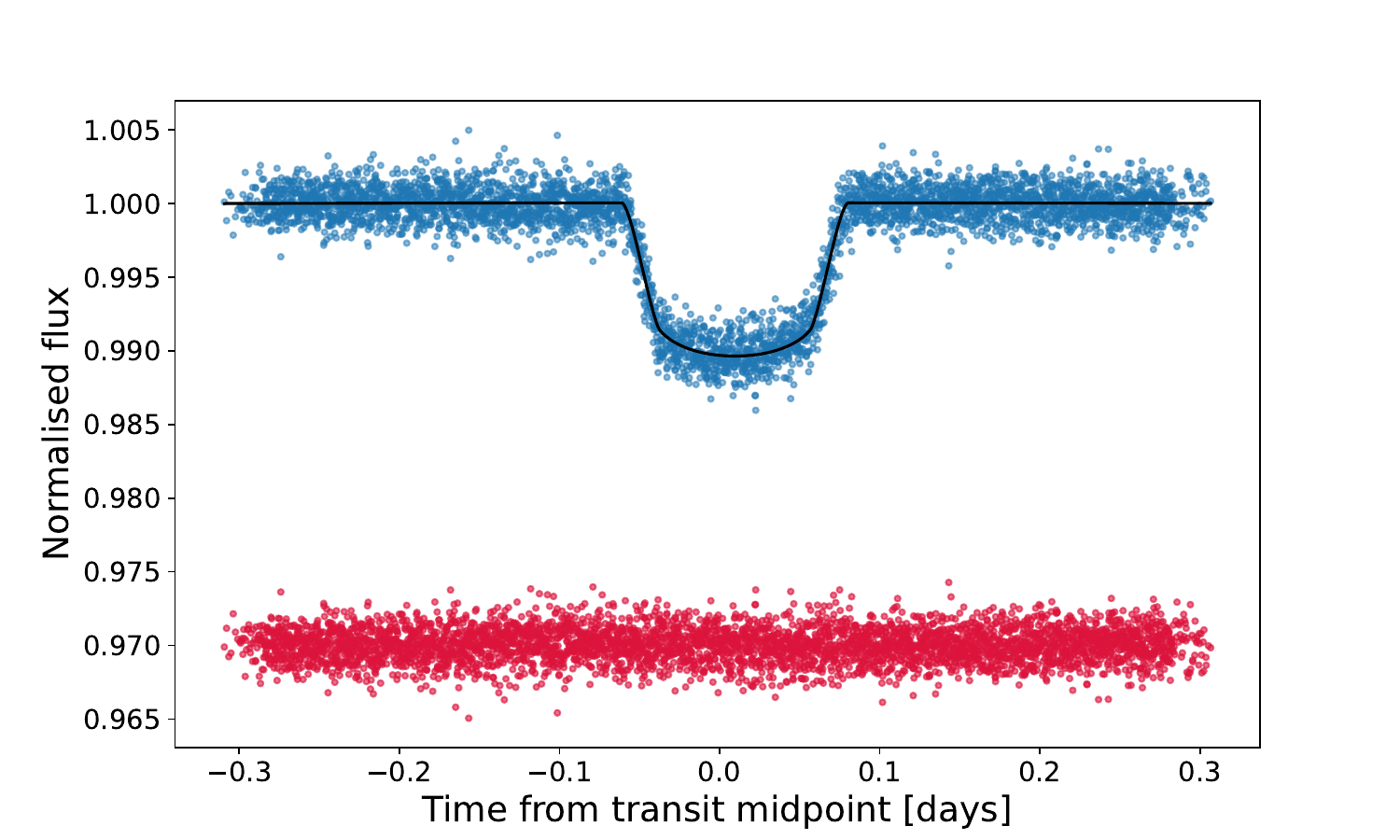}
\end{center}
\includegraphics[height=15cm]{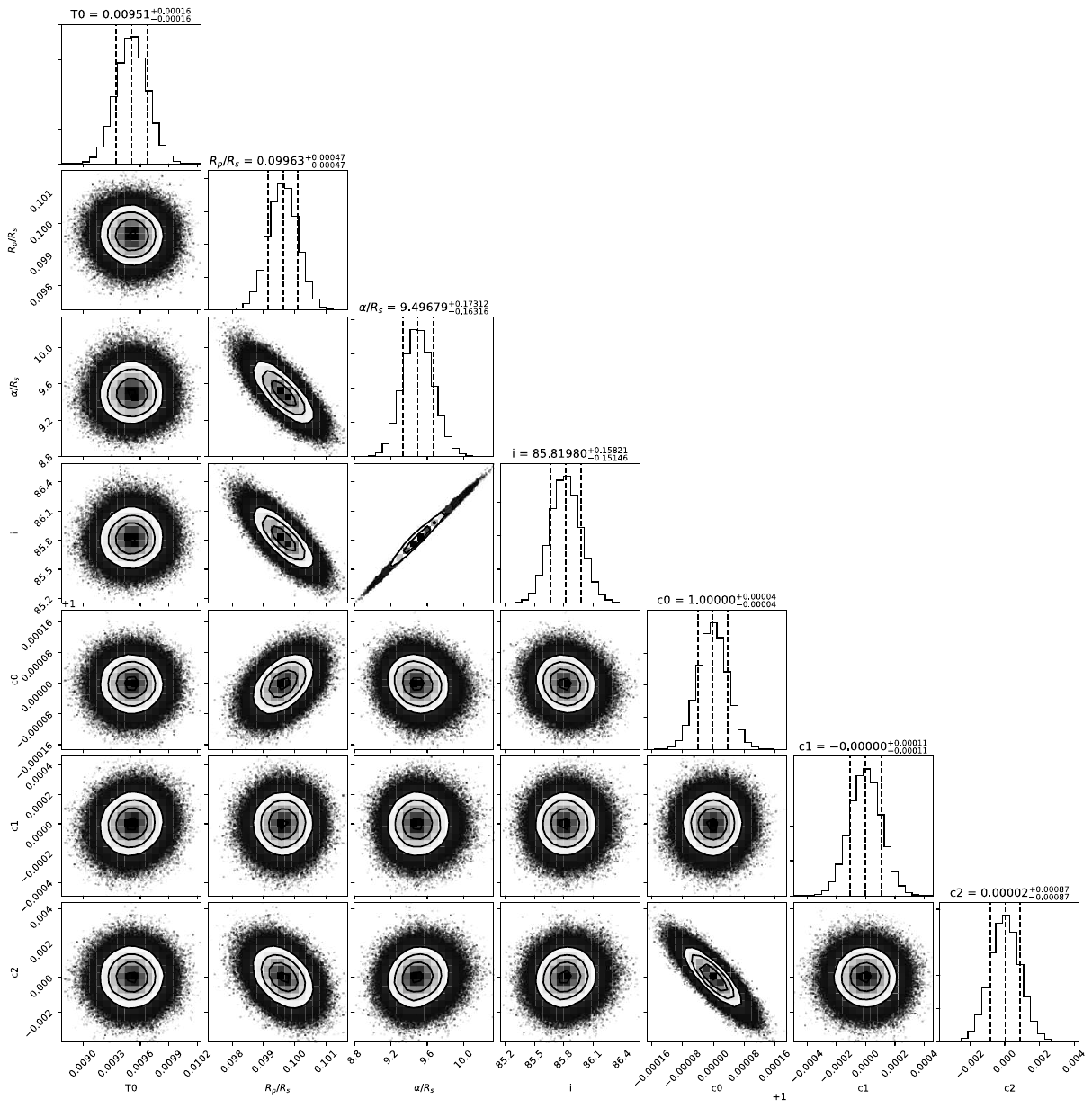}
\caption{The same as Fig. \ref{wasp140b_corner}, but for the exoplanet \mbox{WASP-20 b}.}
\label{wasp20b_corner}
\end{figure*}

\begin{figure*}[ht]
\begin{center}
\includegraphics[height=6cm]{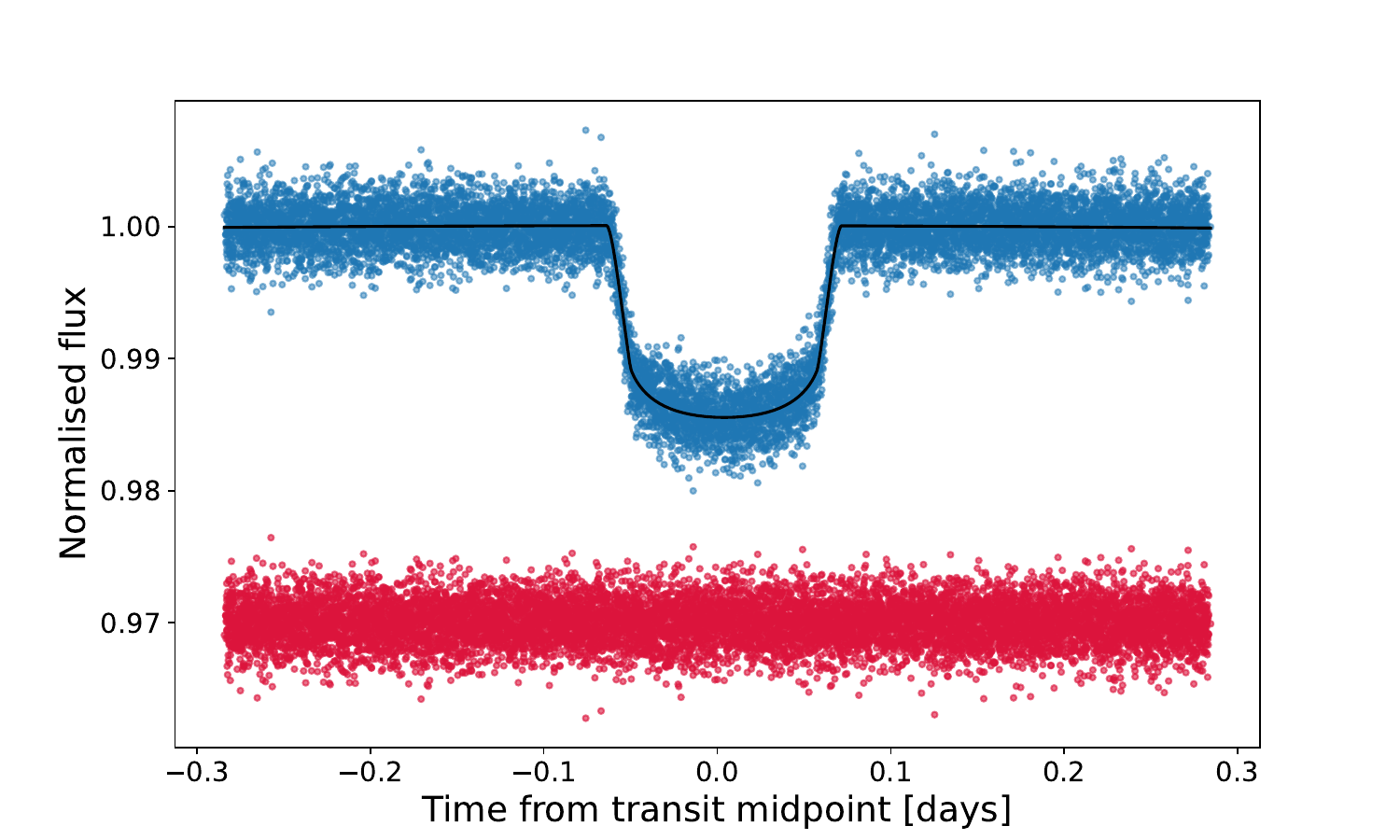}
\end{center}
\includegraphics[height=15cm]{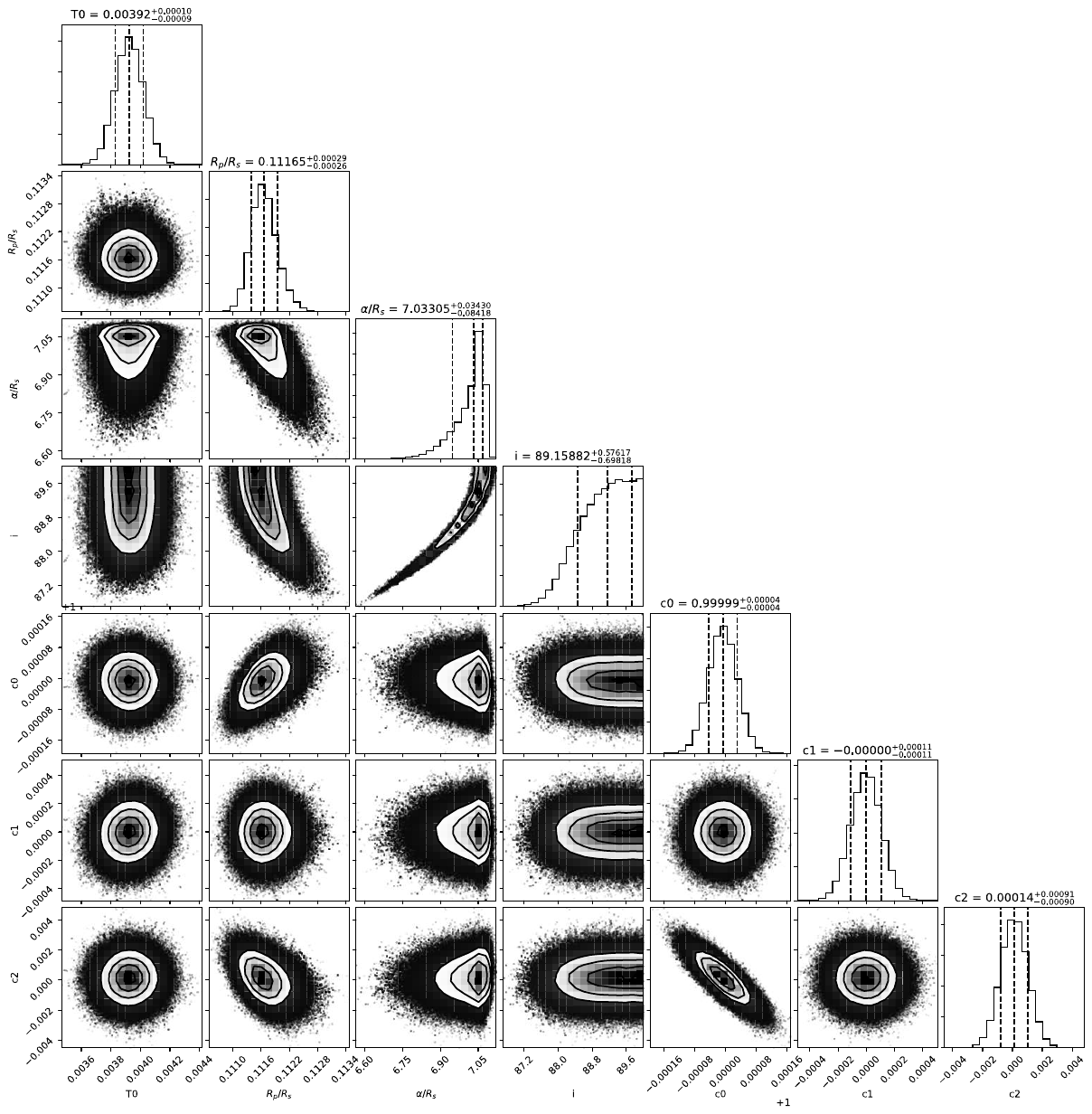}
\caption{The same as Fig. \ref{wasp140b_corner}, but for the exoplanet \mbox{WASP-108 b}.}
\label{wasp108b_corner}
\end{figure*}

\end{document}